\documentclass{aa}  

\usepackage{graphicx}
\usepackage[dvipsnames]{xcolor}
\usepackage{float}
\usepackage[utf8]{inputenc}
\usepackage{tabularx} 
\usepackage{booktabs} 
\usepackage{txfonts}
\usepackage{hyperref}
\hypersetup{colorlinks,allcolors=black}

\begin{document}

\title{The large-scale kinematics of young stars in the Milky Way disc: first results from SDSS-V}

     \author{Eleonora Zari \inst{\ref{1}} \and 
     Jaime Villase\~nor \inst{\ref{2}} \and 
     Marina Kounkel \inst{ \ref{3}} \and  
     Hans-Walter Rix \inst{\ref{2}} \and
     Neige Frankel \inst{\ref{4}, \ref{5}} \and
     Andrew Tkachenko \inst{\ref{6}} \and
     Sergey Khoperskov \inst{\ref{7}} \and
     Elena D'Onghia \inst{ \ref{8}} \and
     Alexandre Roman-Lopes \inst{\ref{9}} \and 
     Carlos Román-Zúñiga \inst{\ref{10}} \and
     Guy S. Stringfellow \inst{\ref{11}} \and
     Jonathan C. Tan \inst{\ref{12}, \ref{13}} \and
     Aida Wofford \inst{\ref{14}} \and
     Dmitry Bizyaev \inst{\ref{15}, \ref{16}} \and
     John Donor \inst{\ref{17}} \and
     Jos{\'e} G. Fern{\'a}ndez-Trincado \inst{\ref{18}}
     Sean Morrison \inst{\ref{19}} \and
     Kaike Pan \inst{\ref{15}} \and
     Sebastian F. Sanchez \inst{\ref{20}} \and
     Andrew Saydjari \inst{\ref{21}}}
     \institute
     {Dipartimento di Fisica e Astronomia, Universit{\`a} degli Studi di Firenze, Via G. Sansone 1, I-50019, Sesto F.no (Firenze), Italy\label{1} \and 
     Max-Planck-Institut f{\"u}r Astronomie, Konigstuhl 17, D-69117,
    Heidelberg, Germany\label{2} \and
      Department of Physics and Astronomy, University of North Florida, 1 UNF Dr, Jacksonville, FL 32224, USA \label{3} \and
     Canadian Institute for Theoretical Astrophysics, University of Toronto, 60 St. George Street, Toronto, ON M5S 3H8, Canada\label{4} \and
      David A. Dunlap Department of Astronomy and Astrophysics, University of Toronto, 50 St. George Street, Toronto, ON M5S 3H4, Canada\label{5} \and
      Institute of Astronomy, KU Leuven, Celestijnenlaan 200D, 3001 Leuven, Belgium\label{6} \and
      Leibniz-Institut f{\"u}r Astrophysik Potsdam (AIP), An der Sternwarte 16, 14482 Potsdam, Germany \label{7} \and
      University of Wisconsin–Madison, Department of Astronomy, Madison, WI, 53706, USA \label{8} \and
      Departamento de Astronom{\'i}a, Facultad de Ciencias, Universidad de La Serena, Av. Raul Bitran 1302, La Serena, Chile \label{9}   
      \and
      Universidad Nacional Aut{\'o}noma de M{\'e}xico, Instituto de Astronom{\'i}a, AP 106, Ensenada 22800, BC, M{\'e}xico \label{10} \and
      University of Colorado Boulder, Boulder CO 80309 USA \label{11} \and
      Department of Astronomy, University of Virginia, Charlottesville, Virginia 22904, USA \label{12} \and
      Dept. of Space, Earth \& Environment, Chalmers University of Technology, SE-41293 Gothenburg, Sweden \label{13}  \and
      Instituto de Astronom{\'i}a, Universidad Nacional Aut{\'o}noma de M{\'e}xico, Unidad Acad{\'e}mica en Ensenada, Km 103 Carr. Tijuana-Ensenada, Ensenada 22860, M{\'e}xico\label{14} \and
      Apache Point Observatory and New Mexico State
      University, P.O. Box 59, Sunspot, NM, 88349-0059, USA \label{15}\and
      Sternberg Astronomical Institute, Moscow State
      University, Moscow, 119234, Russia\label{16} \and
      Department of Physics and Astronomy, Texas Christian University, Fort Worth, TX 76129, USA \label{17} \and
      Instituto de Astronom{\'i}a, Universidad Cat{\'o}lica del Norte, Av. Angamos 0610, Antofagasta, Chile \label{18} \and
      Department of Astronomy, University of Illinois at Urbana-Champaign, Urbana, IL 61801, USA\label{19} \and
      Instituto de Astronom{\'i}a, Universidad Nacional Aut{\'o}noma de M{\'e}xico, A.P. 70-264, 04510, Mexico, D.F., M{\'e}xico \label{20}\and
      Department of Astrophysical Sciences, Princeton University, Princeton, NJ 08544, USA\label{21} 
     }


\abstract{We present a first large-scale kinematic map of $\sim$50,000 young OB stars ($T_{\rm eff} \geq 10,000$ K), based on BOSS spectroscopy from the Milky Way Mapper OB program in the ongoing Sloan Digital Sky Survey V (SDSS-V). Using photogeometric distances, line-of-sight velocities and Gaia DR3 proper motions, we map 3D Galactocentric velocities across the Galactic plane to $\sim$5 kpc from the Sun, with a focus on radial motions ($v_R$). Our results reveal mean radial motion with amplitudes of $\pm 30$ km/s that are coherent on kiloparsec scales, alternating between inward and outward motions. These $\bar{v}_R$ amplitudes are considerably higher than those observed for older, red giant populations. These kinematic patterns show only a weak correlation with spiral arm over-densities. Age estimates, derived from MIST isochrones, indicate that 85\% of the sample is younger than $\sim300$~Myr and that the youngest stars ($\le 30$~Myr) align well with density enhancements. The age-dependent $\bar{v}_R$ in Auriga makes it plausible that younger stars exhibits different velocity variations than older giants. The origin of the radial velocity features remains uncertain, and may result from a combination of factors,  including spiral arm dynamics, the Galactic bar, resonant interactions, or phase mixing following a perturbation. The present analysis is based on approximately one-third of the full target sample. The completed survey will enable a more comprehensive investigation of these features and a detailed dynamical interpretation.}

\date{Received -; accepted -}
\titlerunning{SDSS-V OB kinematics}
\authorrunning{Zari et al.}
\maketitle

\section{Introduction}
The Galactic disc exhibits significant non-axisymmetric structures that fundamentally affect both stellar distributions and kinematics. Spiral arms and the Galactic bar constitute primary drivers of this non-axisymmetry, playing a crucial role in star formation processes and determining the morphological structure of the Milky Way \citep[][]{Antoja2018, Frankel2018, Reid2019, Eilers2020, Poggio2021}.

Recent stellar density mapping efforts have successfully traced segments of the Galactic spiral arms through the spatial distribution of stars. Multiple studies have utilised samples of Upper Main Sequence (UMS) and OBA-type stars \citep{Zari2021, Poggio2021, GaiaDrimmel2023, Xu2023, Ge2024}, A-type stars \citep{Ardevol2023}, and red clump (RC) stars \citep{Lin2022}. These analyses have consistently identified over-densities corresponding to the Perseus and Local arms, as well as features associated with the Sagittarius-Scutum arm complex.

Kinematic analyses have revealed correlations between stellar velocity fields and spiral structure. \cite{GaiaDrimmel2023} demonstrated significant relationships between spiral-like over-densities and mean velocity fields in three dimensions (radial, azimuthal, and vertical) for their OB star sample within $|z| < 0.3$ kpc and $d \leq $2 kpc. Their extended analysis incorporating red giant branch (RGB) stars revealed coherent large-scale velocity patterns, albeit with weaker correlations to spiral structure compared to the OB population. \cite{Eilers2020} identified a large-scale Galactocentric radial velocity feature in their giant star sample, which can be interpreted as the response to a steady-state logarithmic spiral perturbation. \cite{Palicio2023}
identified arc-shaped structures in the radial actions $J_{R}$ of a clean sample of \textit{Gaia} DR3 stars with full astrometric information and line-of-sight velocities available. They suggested that these features could be a response to the spiral arms or the Galactic bar, or could be connected to moving groups.  

With  few exceptions \citep{GaiaDrimmel2023, Poggio2024},  density mapping has mostly relied on ``young"  samples ($\tau_{age}\le t_{orbit}$), while kinematic studies of the Milky Way disc have been based on  ``old" stars ($\tau_{age}\gg t_{orbit}$). This is largely due to a lack of spectroscopic data for large samples of young stars, extending over a significant volume of the Galactic disc.

One of the core programs of the current Sloan Digital Sky Survey incarnation \citep[SDSS-V,][]{Kollmeier2025, Kollmeier2017, Almeida2023, SDSSDR19} has set out to remedy this issue. The Milky Way Mapper OB core program \texttt{mwm\_ob} targets a large ($\sim$0.5 million stars) sample of massive hot stars utilising the Baryon Oscillation Spectroscopic Survey (BOSS) spectrograph, installed at the Apache Point Observatory (APO) in New Mexico, USA \citep{Gunn2006} and the du Pont 2.5 m telescope at Las Campanas Observatory (LCO) in Chile \citep{Bowen1973}. BOSS\footnote{Details of BOSS spectrographs are at
https://www.sdss.org/instruments/boss-spectrographs/} is an optical spectrograph covering a wavelength range of 3622-10354 \r{A} with a resolution of R $\sim$ 1800 \citep{Smee2013}.
\cite{Zari2021} outlined the scientific goals of the \texttt{mwm\_ob} program and presented the selection function of the target sample. 


In this paper we describe the properties of sample observed as of February 2025, and construct a large-scale kinematic map of the Galactic disk for young stars, which provides a novel view on the young Milky Way disc, different from what is seen for older populations.
In Section \ref{sec:data} we summarize the selection function of the SDSS-V Hot Star Sample, focussing on the differences with respect to \cite{Zari2021} and \cite{GaiaDrimmel2023}. In Section \ref{sec:results} we compute the 3D velocities and ages for our sample members, and construct a large-scale map of the Galactocentric radial velocities, $\bar{v}_R$. We discuss our findings in Section \ref{sec:discussion} by comparing them with mock stellar catalogues generated from the \texttt{AURIGA} cosmological simulations and observations of older stellar populations. Finally, in Section \ref{sec:conclusions} we draw our conclusions and discuss future work.

\section{The SDSS-V Hot Star Sample} \label{sec:data}
SDSS-V's science is organized in ``mapper programs" \citep[][]{Kollmeier2025, SDSSDR19}, including the \emph{Milky Way Mapper} program, whose science in turn is parsed operationally in sample ``cartons".
In this Section, we describe the selection function of the \texttt{mwm\_ob} target carton and the status of the observations as of the 2nd of February 2025 (start of survey thru MJD=60708). We chose this date because it will  be the cut for SDSS data release 20 (DR20), which is scheduled no earlier than Summer 2026. We present the criteria that we used to remove cooler sources from the initial \texttt{mwm\_ob} targets after the spectra had been obtained, and the spectral parameters and line of sight velocities used.

\subsection{Selection function of the SDSS-V \texttt{mwm\_ob} Targets}
\begin{figure*}[!ht]
    \includegraphics[width = \hsize]{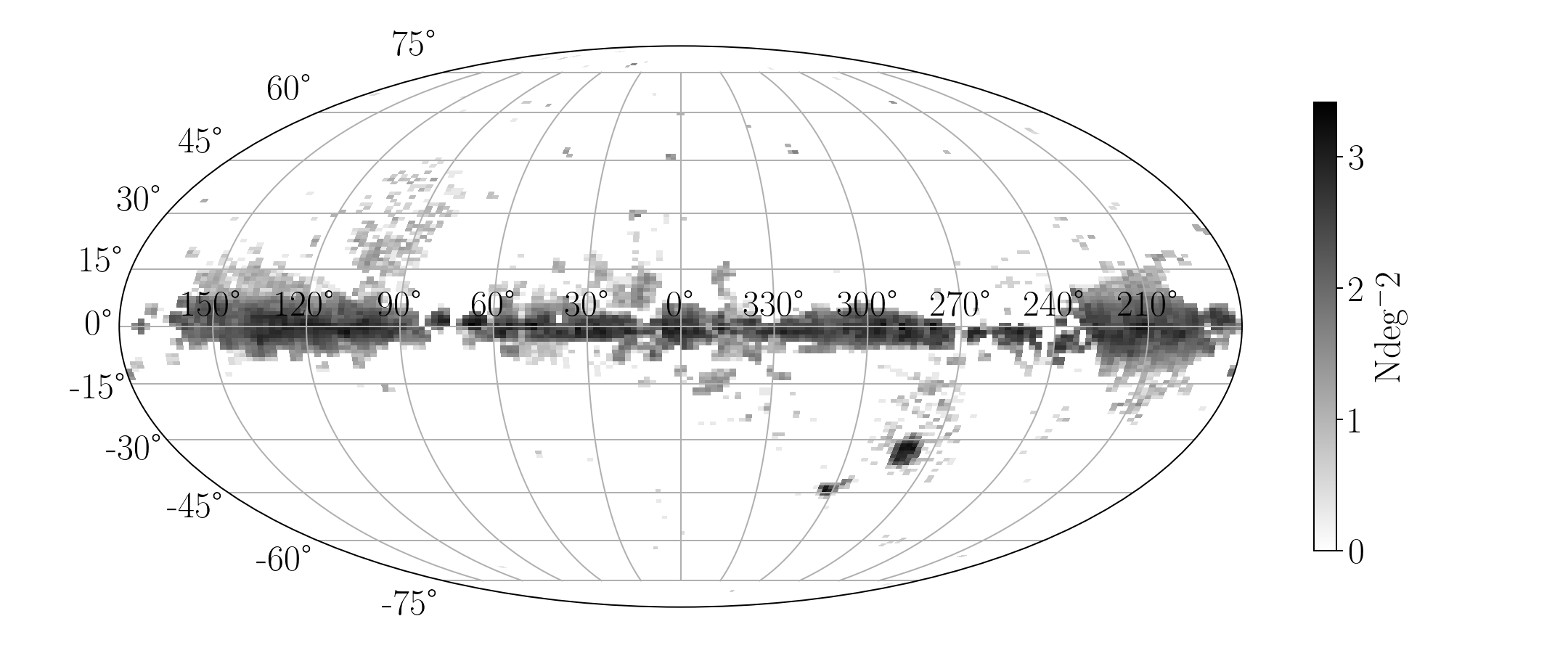}
    \includegraphics[width = \hsize]{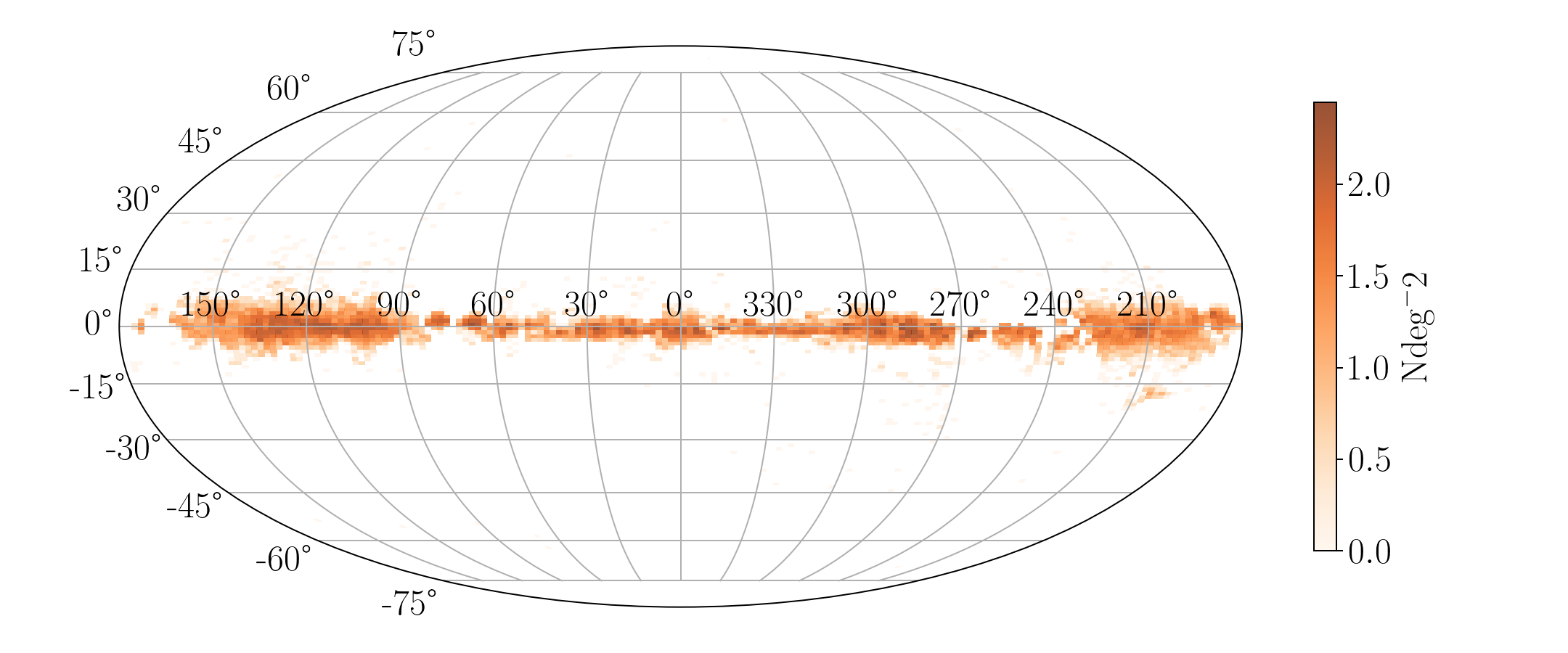}
    \caption{Distribution in the sky (Galactic coordinates) of the sources of the SDSS-V \texttt{mwm\_ob} program observed until Feb 2025 (top, gray, 165,460 sources) and of the filtered sample described in Section \ref{sec:filtered} (bottom, orange, around 45,487 stars). The sources of the \texttt{mwm\_ob} program are mostly located in the Galactic plane and the MCs. }
    \label{fig:sky_distribution}
\end{figure*}
The targeting strategy of the \texttt{mwm\_ob} sample in SDSS-V is presented in \citet[][Z21]{Zari2021}. In Z21, the target selection strategy prioritized completeness over purity, with the understanding that SDSS-V spectroscopic data would later allow us to refine the sample and improve its purity. The target selection  relies on a combination of photometric and astrometric criteria, based on \textit{Gaia} E/DR3 and 2MASS \citep{Skrutskie2006} astrometry and photometry. 
Z21's target sample (described in their Section 2.1) consisted of  stars more luminous than $M_K = 0$~mag (where $M_K$ is the absolute magnitude in the 2MASS $K_s$ band and $\varpi$ indicates the parallax in mas), or more precisely,
\begin{equation}
    \varpi < 10^{(10 - Ks)/5}.
\end{equation}
In practice, the SDSS-V survey strategy \citep["Robostrategy",][]{Blanton2025} -- and in particular the need to reduce the size of our sample from around 1 Ml to $\sim$ 0.5 Ml sources -- required a slightly tighter selection:
\begin{equation}
    \varpi < 10^{(10 - Ks- 0.6)/5},
\end{equation}
which corresponds to $M_K < -0.6$~mag. The latter condition ideally selects stars of spectral type earlier than B3V \citep[see e.g.][]{Pecaut2013}. Between the start of the survey and the 2nd of  February 2025 (MJD 60708), SDSS-V obtained  459,679 spectra of  165,460 distinct targets in the $\texttt{mwm\_ob}$ program. 
Figure \ref{fig:sky_distribution} (top) shows the distribution of sources in the sky (Galactic coordinates). Whilst the coverage is not uniform, much of the Galactic plane has been already observed, which allows for a first exploration of the global kinematic properties of the sample.

We compared our sample with the \citet[][D23]{GaiaDrimmel2023} sample of O- and B-type stars. D23 (see Appendix \ref{appendix:comparisonDrimmel})  selected sources based on their effective temperatures $T_\mathrm{eff, GSPPHOT}$ and $T_\mathrm{eff, ESP-HS}$. 
The GSPPHOT temperature estimates are derived by combining $\textit{Gaia}$ GBP/GRP spectrophotometry, astrometry, and G band photometry. 
The ESP-HS estimates  are derived from a software module (ESP-HS) that was optimised specifically for hot stars and uses the BP/RP spectrophotometry, without the astrometry, together with the RVS spectra if they are available as well \citep[][]{Creevey2023, Fouesneau2023}. For stars with only GSP-Phot temperatures, D23 used two conditions: $T_\mathrm{eff, GSPPHOT} > 10,000$~K and the  ESP-HS spectral type flag set to O, B, or A; for stars with only ESP-HS temperatures, they required that the effective temperature be in the range 10,000 $< T_{\mathrm{eff, ESPHS}} < $ 50,000 K; for sources with both sets of stellar parameters, they required the effective temperature to be $T_{\mathrm{eff, GSP-Phot}} > $ 8,000 K and $T_\mathrm{eff, ESPHSH} >$ 10,000 K. These selection criteria resulted in a list of 923,700 sources.

Around 25\% of the \texttt{mwm\_ob} targets are in common with D23. 
The majority of the sources (90\%) that are included in D23
but are not in the \texttt{mwm\_ob} program are between  $-0.6 < M_\mathrm{K} < 2$~mag, 
and are thus excluded by our more stringent selection cuts (see Appendix \ref{appendix:comparisonDrimmel}). These could be either later B-type/early A-type stars or heavily extincted sources.

On the other hand, $\sim$75\% of the sources in \texttt{mwm\_ob} were not selected by D23. 
These stars are either cooler contaminants of the \texttt{mwm\_ob} program (see Sec. 2.3 in Z21, and the next Section) or genuine hot stars, whose spectral parameters have not been published in \textit{Gaia} DR3. We filter out the cooler contaminants in the next Subsection. 

\subsection{The Hot Star Sample}\label{sec:filtered}
As mentioned above, the photometrically-selected \texttt{mwm\_ob} targets include a significant fraction of A-type stars and a modest fraction of cooler contaminants (see Sec. 2.3 in Z21). We therefore filtered the \texttt{mwm\_ob} sample for the subsequent analysis, to obtain a sample of hot stars with well determined spectroscopic parameters and distances (Hot Star Sample, HSS). The filtering is  based on four parameters:
signal-to-noise ratio of the observed single epoch spectra, parallax error, RUWE \citep[renormalised unit weight error, ][]{Lindegren2021}, and the effective temperature derived from the BOSS spectra. 

\begin{figure}
    \centering
    \includegraphics[width = 0.5\textwidth]{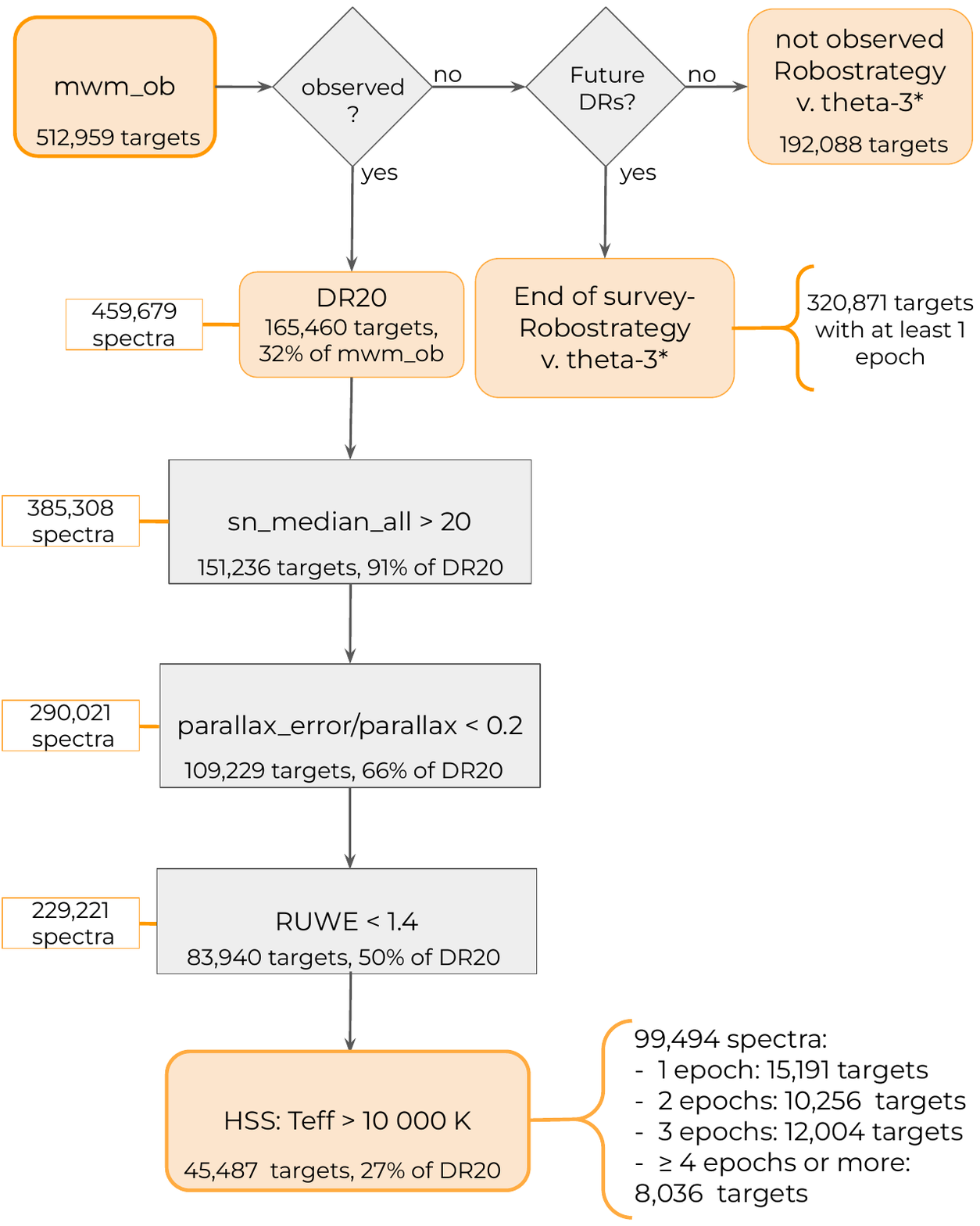}
    \caption{Flowchart illustrating the steps to construct the HSS described in Sec. \ref{sec:filtered}. The notation "v.theta-3*" indicates that the numbers are provisional and refer to the specific Robostrategy version named "theta-3" \citep[][]{Blanton2025}. The survey efficiency has improved and the final number of observed targets will likely increase in the future.}
    \label{fig:flow_chart}
\end{figure}

Figure \ref{fig:flow_chart} shows a flowchart detailing the steps to construct the HSS. DR20 contains around a third of the target sample of the \texttt{mwm\_ob} program. By the end of the survey, around 60\% of the sources will have at least one observation. We note however that this is an estimate based on the "theta-3" version of Robostrategy \citep[][]{Blanton2025}. The survey strategy has slightly changed and the survey efficiency has improved, therefore the final numbers will be different.

We required  the median signal-to-noise ratio (\texttt{sn\_median\_all}) of the single epoch spectra  in the BOSS pipeline summary file (reduction version \texttt{v6\_2\_0}, Morrison S. et al., \emph{in prep.}) to be SNR $>$ 20, to ensure the reliability of the estimated spectral parameters. 

We selected sources with parallax relative error $\sigma_{\varpi}/\varpi < 0.2$ and RUWE < 1.4. The former cut aims at selecting sources with good distance determinations, the latter seeks to remove sources with spurious parallaxes or proper motions \citep[and possibly binaries, ][]{Belokurov2020}.

\begin{figure}
    \includegraphics[width = 0.5\textwidth]{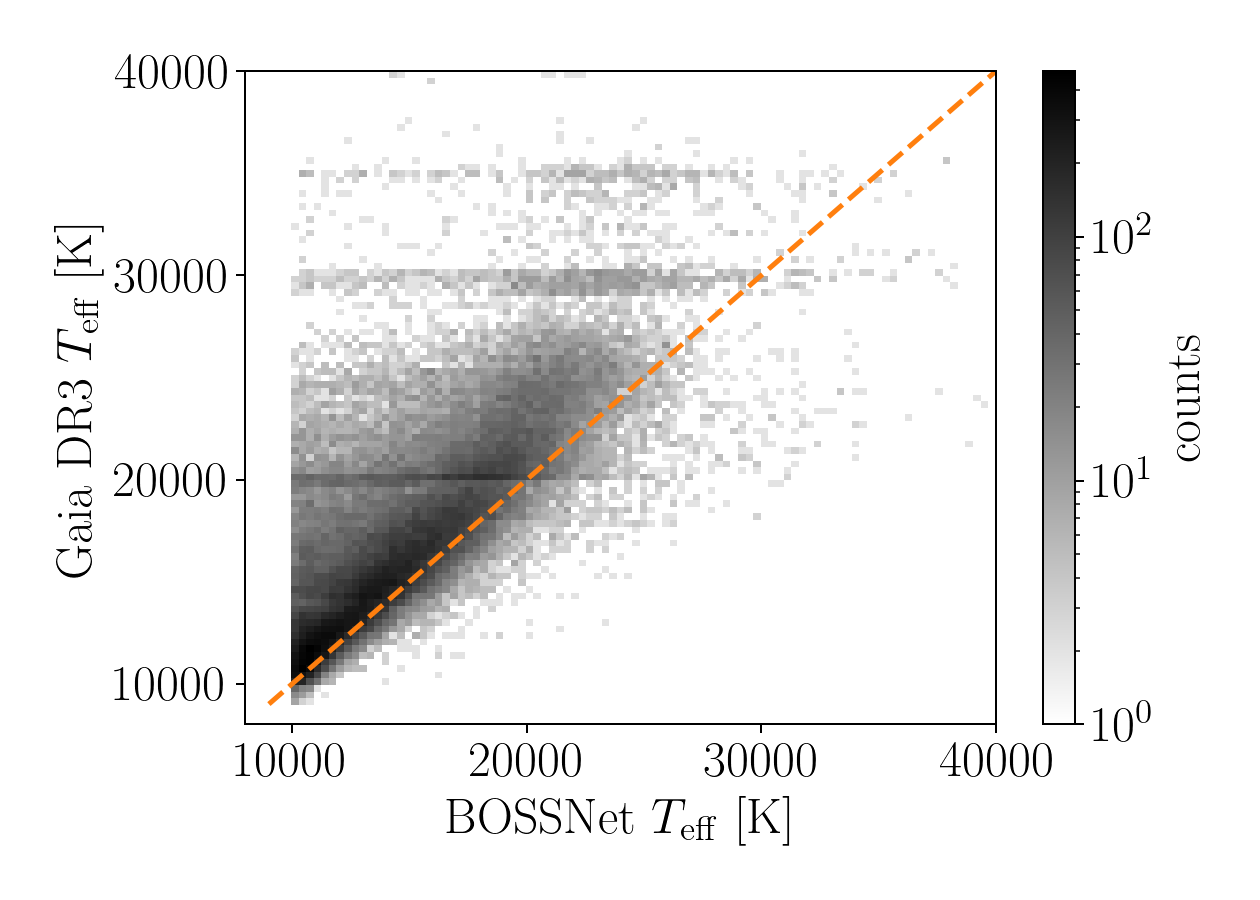} \\
     \includegraphics[width = 0.5\textwidth]{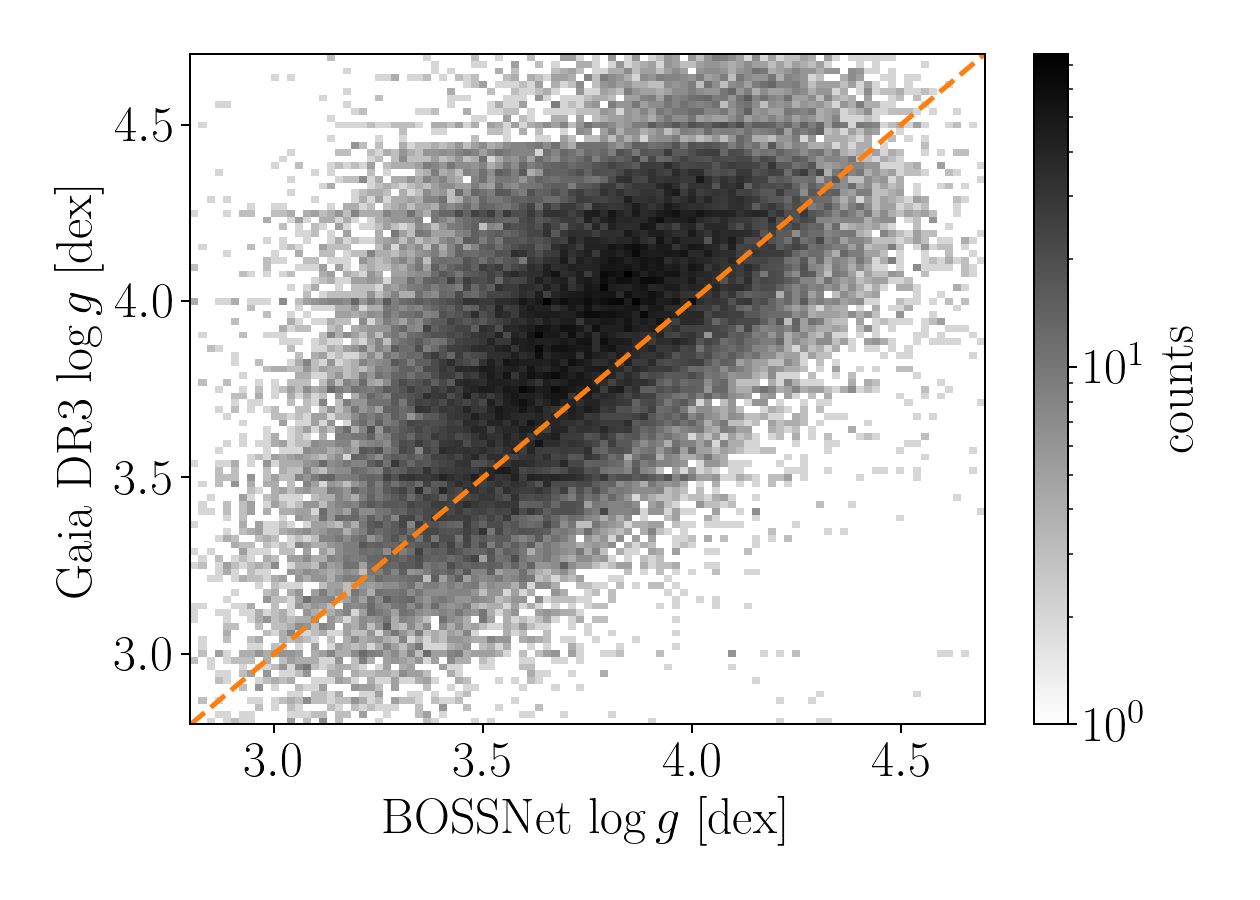}
     \caption{Comparison between the BOSS Net parameters and \textit{Gaia} DR3 ESP-HS parameters for the Hot Stars Sample. The dashed orange lines correspond to the 1:1 relation.}
    \label{fig:compare_params}
\end{figure}

We  selected stars with  $T_{\mathrm{eff}}>$ 10,000~K, according to the BOSS Net estimates \citep{Sizemore2024}. BOSS Net is a neural network that predicts self-consistent $T_{\mathrm{eff}}$ and $\log g$ (with respective precision of 0.008
and 0.1 dex at S/N $\sim$ 15) in optical (BOSS and LAMOST) spectra for stars of all spectral types. For O-, B-, and A-type stars it was trained  on LAMOST spectra with parameters determined by the  HotPayne \citep[][]{Xiang2022}.  Additionally, it provides
an independent estimate of stellar line-of-sight velocities ($v_{l.o.s., \mathrm{BNet}}$).  
Figure \ref{fig:compare_params} shows a comparison of spectral parameters derived by BOSS Net (left $T_{\mathrm{eff}}$, centre $\log g$) with those estimated by the \textit{Gaia} ESP-HS pipeline. Both $T_{\mathrm{eff}}$ and $\log g$ derived by the ESP-HS pipeline are on average higher than those provided by BOSS Net, but broadly agree with one another. Around  70\% and 80\% of sources have parameters within 2$\sigma$, for $T_{\mathrm{eff}}$ and $\log g$ respectively, which is sufficient for the purposes of this study. We refer to Villase\~nor et al. (\emph{in prep.}) for a thorough validation of the spectral parameters with synthetic spectra. 

Figure \ref{fig:sky_distribution}(bottom) shows the distribution in the sky of the HSS. The selection criteria of the HSS effectively select sources in the Galactic Plane, excluding  high-latitude sources and the Milky Clouds (MCs) \footnote{Although there is no general consensus in the community, we decided to adopt this nomenclature following \cite{delosreyes2023PhyOJ..16..152}}. The MCs sources are mostly excluded by the requirement $\sigma_{\varpi}/\varpi < 0.2$. The HSS  results in a sample of 99,494 spectra of 45,487 stars,  with minimum and maximum distance respectively of 46 pc and 11 kpc, with average distance error  $\sim 20, 70, 150, 300, 500$~pc at 1, 2, 3, 4, 5~kpc.  With respect to the D23 sample, the SDSS-V Hot Star filtered sample covers  a wider area of the Galactic plane. This is illustrated and further discussed in Appendix \ref{appendix:comparisonDrimmel}.

Figure \ref{fig:logg-teff} shows the distribution of targets of the \texttt{mwm\_ob} program passing the quality cuts described above (gray histogram) and the HSS (orange histogram) in the $\log T_{\mathrm{eff}} - \log \mathrm{g}$ diagram. The main contaminants of the \texttt{mwm\_ob} program are A-type and F-type stars, respectively 33,309 and 4,065 sources, that are difficult to separate from O- and B-type stars by using only \textit{Gaia} and 2MASS photometry. There is also a small number of giants at $\log g \sim 2.5$~dex. These sources are likely included because they have the same colours in the $G-K_s$ vs. $J-H$ colour-colour diagram as reddened O- and B-type stars (see Fig. 3 in Z21). Among the O- and B-type stars, 29,503 sources have temperature between 10,400 K and 17,000 K (spectral type between B9 and B3); 10,632 sources have temperature between 17,000 K and 31,9000 K (B3-O9.5); 213 sources have temperature higher than 31,900 K (O9.5 and above), from which 131 are classified as O stars in SIMBAD.

\begin{figure}
    \centering
 \includegraphics[width = \linewidth]{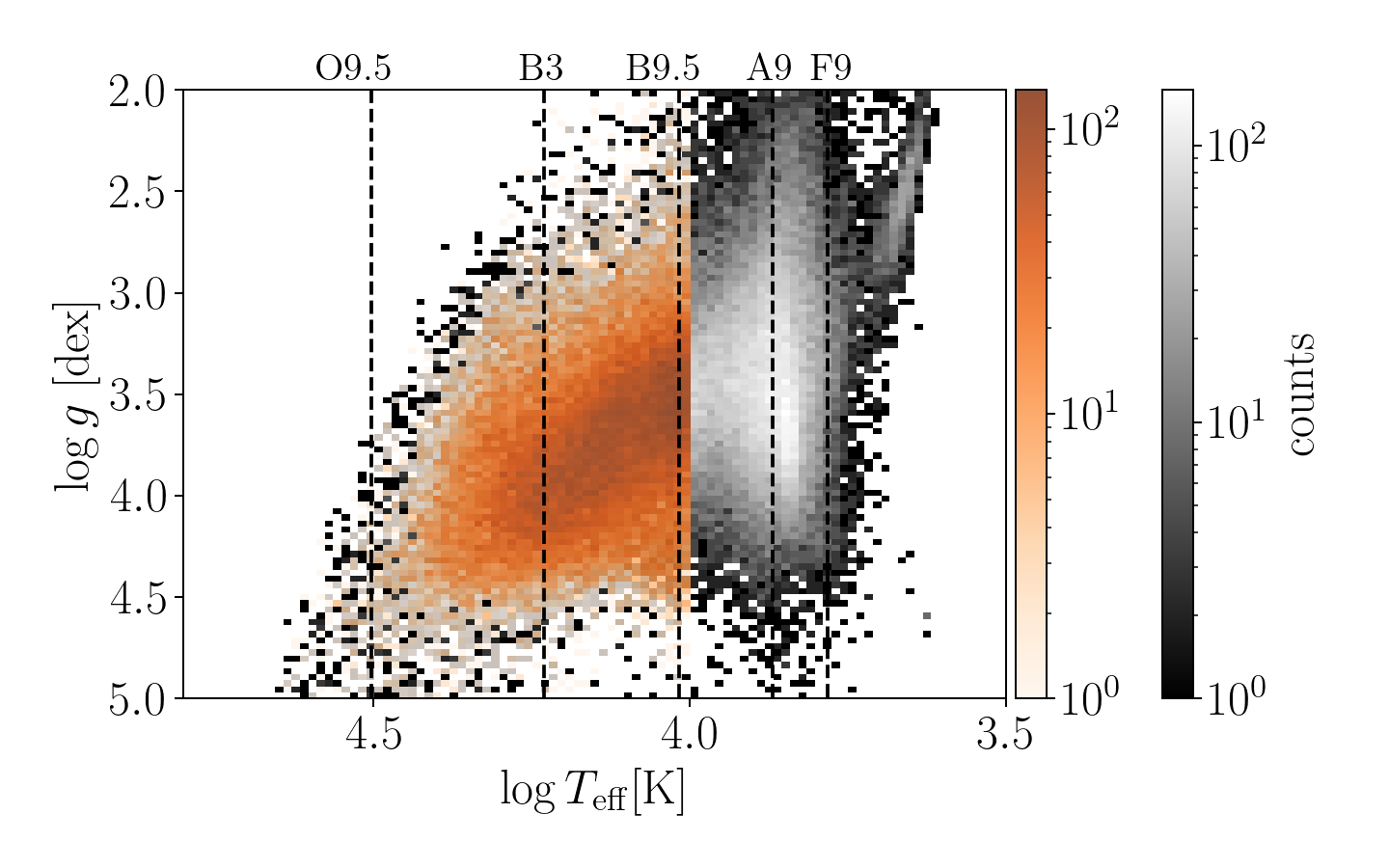}     \caption{Kiel diagram of the sources in the \texttt{mwm\_ob} program (gray histogram) passing the quality cuts described in Section \ref{sec:filtered} and of those in the HSS (coloured histogram). The dashed vertical lines mark the following temperatures: 31,000 K (O9.5V type star), 17,000 (B3V type star), 10,400 K (B9V type star), 7,400 K (A9V type star), and 6,050 K (F9V type star).}
    \label{fig:logg-teff}
\end{figure}

We compared the HSS with the ALS-III catalogue \citep{pantaleoni2025}, which consists of a reliable sample of  Galactic OB stars. We downloaded the ALS-III catalogue from the web interface  \url{https://als.cab.inta-csic.es} (20,387 sources), and selected the sources flagged as "M". This sample (14,273 sources) contains stars that are above the ZAMS and the 20kK extinction track. In addition, these stars have \textit{Gaia} DR3 photometry consistent with being massive stars, and have been reported as O- or B-type stars in the literature or have been spectroscopically validated. We cross-matched the ALS-III "full" and  "M" sample with: 1) the HSS; 2) the end of survey sample (theta-3 version of Robostrategy); 3) the target sample described in Z21 with the condition $M_K< -0.6$~mag; 4) the full target sample presented in Z21. We performed positional cross-matches, with a radius of 1". Table \ref{tab:ob_crossmatches} reports the results of the cross-matches.\begin{table*}[h!]
\centering
\caption{Cross-match with the ALS-III catalogue \citep[][]{pantaleoni2025}. }
\label{tab:ob_crossmatches}
\begin{tabular}{l|cccc}
\toprule
Catalogue & HSS & Theta-3 ($\ge$1 epoch) & Zari+21, $M_K<-0.6$ & Zari+21 (full)\\
\footnotesize\it (N total) &
\footnotesize\it N = 45,487 &
\footnotesize\it N = 320,871 &
\footnotesize\it N = 512,959 &
\footnotesize\it N = 988,202 \\
\midrule
\shortstack[l]{ALS-III ``M''\\\footnotesize\it N = 14,273}   & 2,960 & 6,490 & 11,525 & 12,503 \\
\shortstack[l]{ALS-III (full)\\\footnotesize\it N = 20,387} & 3,005 & 7,656 & 13,748 & 15,799 \\
\bottomrule
\end{tabular}
\end{table*}

The number of stars in common between the HSS and the ALS-III catalogues is currently relatively low (around 20\% and 15\% of the ALS-III "M" and "full" samples) but will increase to around the 50\% (40\%) of the ALS-III "M" ("full") targets at the end of the survey.  The initial target sample presented in Z21 contained around  87\% (80\%) of the ALS-III "M" ("full") sample. Most ot the ALS-III stars not included in the Z21 target sample  have $M_K > -0.6$~mag. Finally, the remaining 40,129 (40,083) sources in the HSS  not included in the ALS-III "M" ("full") sample include
130  O-type stars ($T_\mathrm{eff} > 31,900$~K) and 8,483 (8,458 when comparing with the "full" sample)  early B-type stars ($T_{\mathrm{eff}} > 17,000$~K). 

\subsection{Line-of-sight velocities}\label{sec:vel}
To derive the 3D kinematic properties of our sample we combine astrometric data from \textit{Gaia} DR3 with SDSS-V's line-of-sight velocities $v_{l.o.s.}$.
In the SDSS-V context \citep[][]{Kollmeier2025, SDSSDR19}, there are two sets of estimates for line-of-sight velocities: those determined by the overall BOSS pipeline $v_{l.o.s, \mathrm{XCSAO}}$ and those determined by the BOSS Net method $v_{l.o.s, \mathrm{BNet}}$. 
The $v_{l.o.s, \mathrm{XCSAO}}$ are determined using the python implementation of the cross-correlation technique first proposed by \cite{Tonry1979}, pyXCSAO \citep[][]{Kounel2022zndo}. 
pyXCSAO  performs Fourier-based cross-correlation between an observed spectrum and template Phoenix spectra  \citep{PhoenixModels2013}. It also estimates the \texttt{XCSAO\_RVX} parameter, which measures the  
strength of the correlation peak relative to the noise.
In our analysis we use $v_{l.o.s., XCSAO}$ for stars with \texttt{XCSAO\_RVX} $>$ 6 (95\% of the targets). Only if the strength of the correlation is lower than this boundary do we use $v_{l.o.s, \mathrm{BNet}}$. 

For targets with multi-epoch observations, we determine $v_{l.o.s.}$ as the average of the available epochs. We note that for most sources there are not enough epochs to compute orbital solutions and, therefore, the systemic velocity of binary sistems that were not excluded by previous cuts. For single stars, this process improves the $v_{l.o.s}$ precision; for binaries (or multiples), the resulting $v_{l.o.s.}$ better approximates the centre of mass motion. We tested this method by simulating a population of stars with 50\% binary fraction, observed at the same MJD's and with the same errors on line of sight velocities as our observed HSS. The results of these  simulations are further discussed in Appendix \ref{appendix:vlos}.


Line-of-sight velocity ($v_{l.o.s.}$ ) uncertainties for single-epoch hot stars are typically  $\sim$10~km/s (for both XCSAO and BOSS Net) and increase slightly with effective temperature. This trend arises because SDSS-V uses a fixed 15-minute exposure time for all sources, resulting in similar SNRs. Further, at higher temperatures, the hydrogen lines that dominate the cross-correlation become broader, reducing velocity precision.
For stars with a single observation, we adopted the uncertainty provided by XCSAO/BOSS Net; for stars with multiple observations, we calculated $\sigma_{v_{l.o.s.}} = std(v_{l.o.s})/\sqrt{N_{obs}}$.  

\begin{figure}
    \centering
    \includegraphics[width = \hsize]{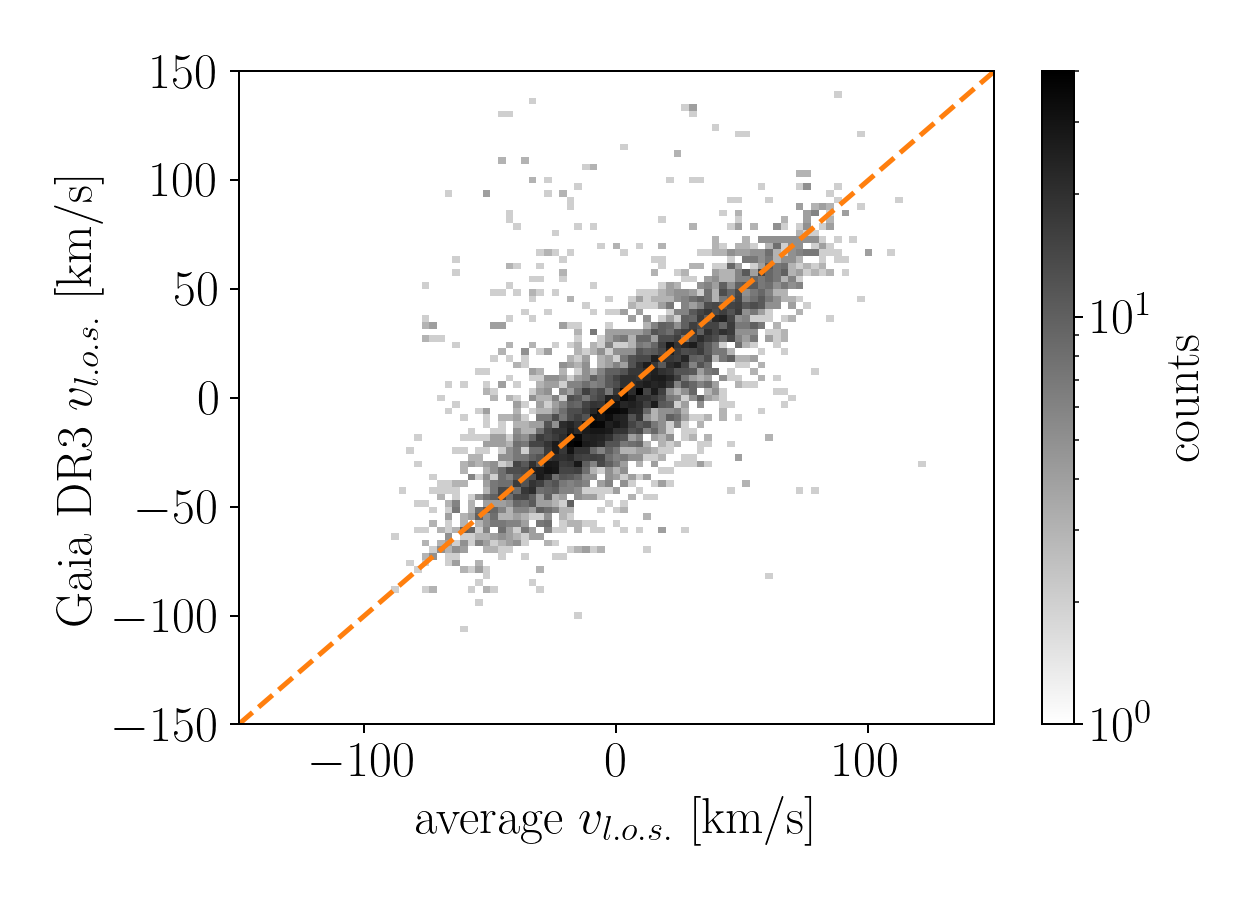}
    \caption{Comparison between \textit{Gaia} DR3 and SDSS-V XCSAO/Boss Net line of sight velocities (in km/s). The orange dashed line corresponds to the 1:1 relation.}
    \label{fig:compare RV}
\end{figure}

To validate the $v_{l.o.s}$ estimates, we compare them with \textit{Gaia} DR3  line-of-sight velocities (Fig. \ref{fig:compare RV}). As prescribed by \cite{Blomme2023},  we applied the following correction to the \textit{Gaia} DR3 $v_{l.o.s.}$ for OB stars:
\begin{equation}
    v_{l.o.s, Gaia} = \mathrm{radial\_velocity - 7.98 + 1.135grvs\_mag.}
\end{equation}
The sub-set of sources for which a \textit{Gaia} DR3 $v_{l.o.s.}$ is available and reliable is relatively small (around 10\% of the initial \cite{GaiaDrimmel2023} query), and the cross-match with our sample is of 5959 single sources. The agreement between these estimates is satisfactory, with around 50\% of the line of sight velocity in agreement within 1$\sigma$ and 75\% in agreement within 2$\sigma$. Outliers can be due to the use of the wrong spectral fitting template by  the \textit{Gaia} and or the XCSAO pipeline, or by strong line-of-sight velocity variations, for instance due to binarity.  


\section{Streaming motions and age maps}\label{sec:results}
In the previous Section we have constructed a well-defined sample of spectroscopically confirmed hot massive stars in the Milky Way disc (the HSS). In this Section we  derive the kinematic properties and the ages of this sample. We  present maps that can be constructed from it, focusing on  the distribution of Galactocentric radial velocities (Fig. \ref{fig:map-galcenrv}) and ages (Fig. \ref{fig:age_plane_all}). 

\subsection{Deriving the 3D positions and kinematics} \label{section:3D pos and kin}
To compute stellar positions in  Cartesian Galactic/Galactocentric coordinates $X,Y$, and $Z$ we adopted the Sun’s Galactocentric position $(R, Z) = (8.2, 0.025)$ kpc \citep{gravity2019, juric2008}  and Galactocentric cylindrical velocities $(v_R, v_{\phi}, v_z )= (11., 232.24, 7.25)$ km/s  \citep[][]{Schonrich2010}, and photogeometric distances for each target \citep{Bailer-Jones2021}. To derive the radial, azimuthal and vertical Galactocentric velocities for each target and the associated uncertainties we followed the procedure described in D23 (see their Sections 3.2 and 3.3). The median Galactocentric radial velocity uncertainty $\bar{\sigma}_{v_R}$ is around 13 km/s for single sources, and around 8 km/s for sources with $N\geq2$ observations. 

\begin{figure}
    \centering
    \includegraphics[width=\linewidth]{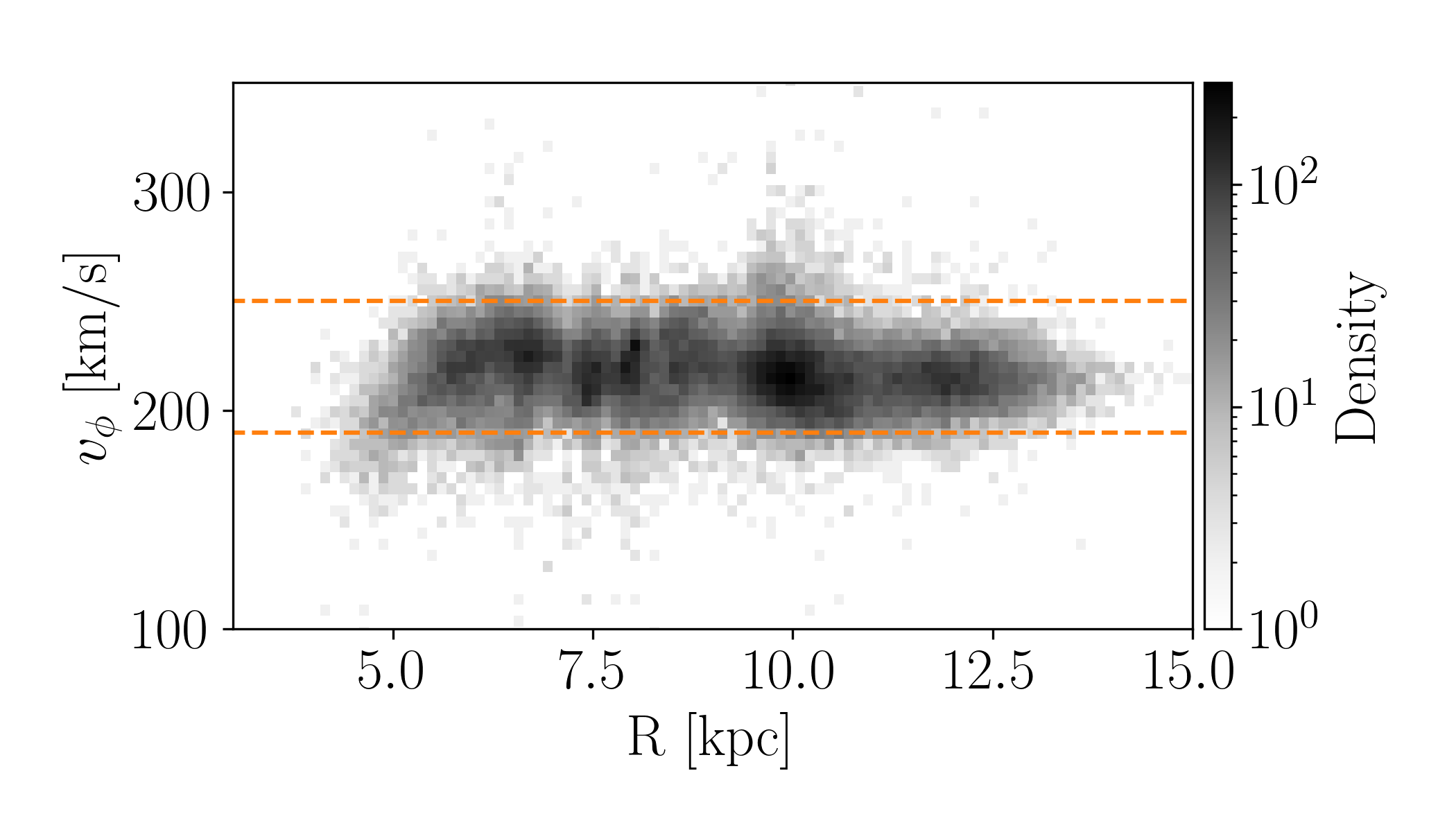}
    \caption{Azimuthal velocity $v_{\phi}$ as a function of Galactocentric radius $R$. The histogram shows the azimuthal velocity distribution. The orange dashed lines indicate the boundaries of our selection, between 190 and 250 km/s.}
    \label{fig:vphi}
\end{figure}

\begin{figure*}
\includegraphics[width = 0.49\hsize]{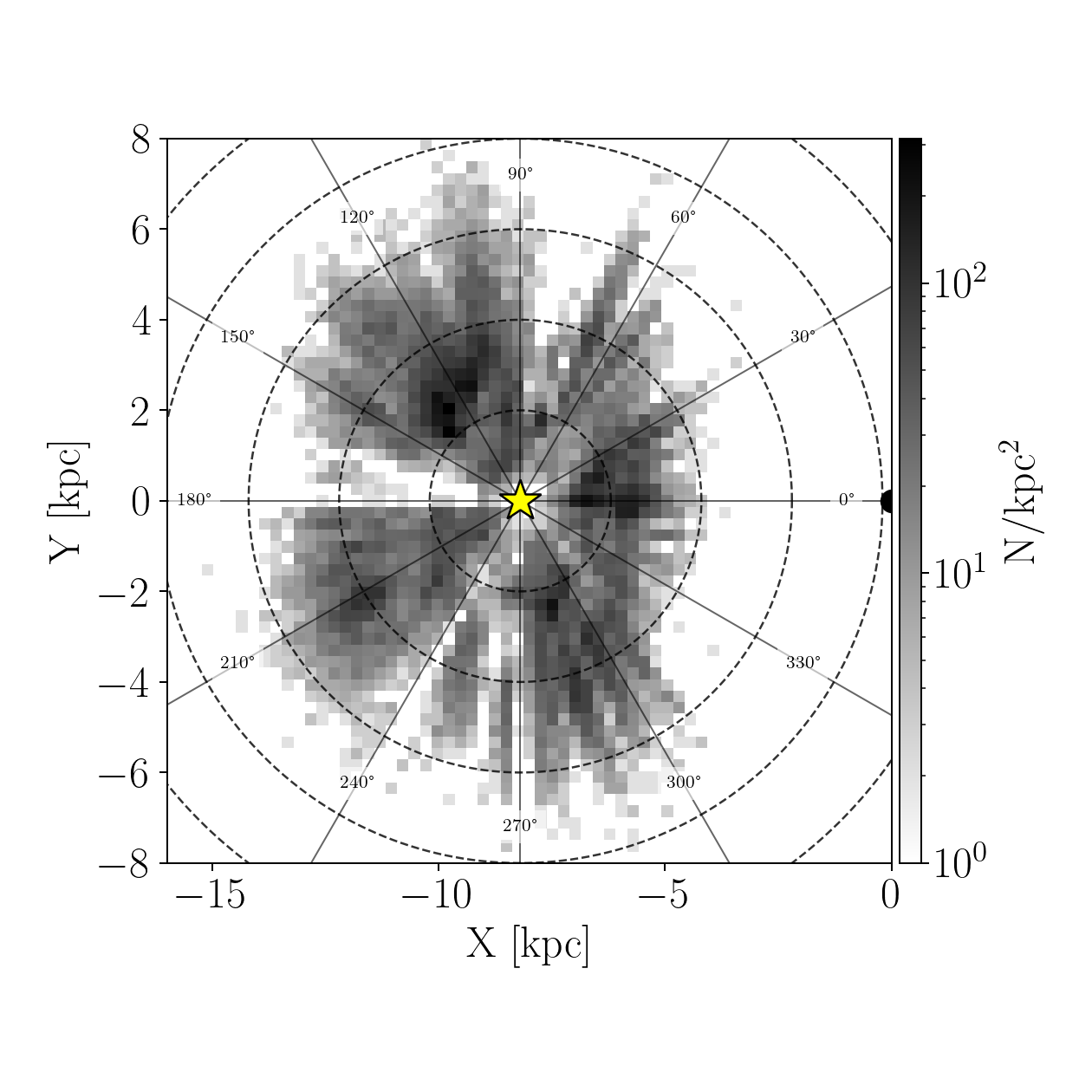}
\includegraphics[width = 0.49\hsize]{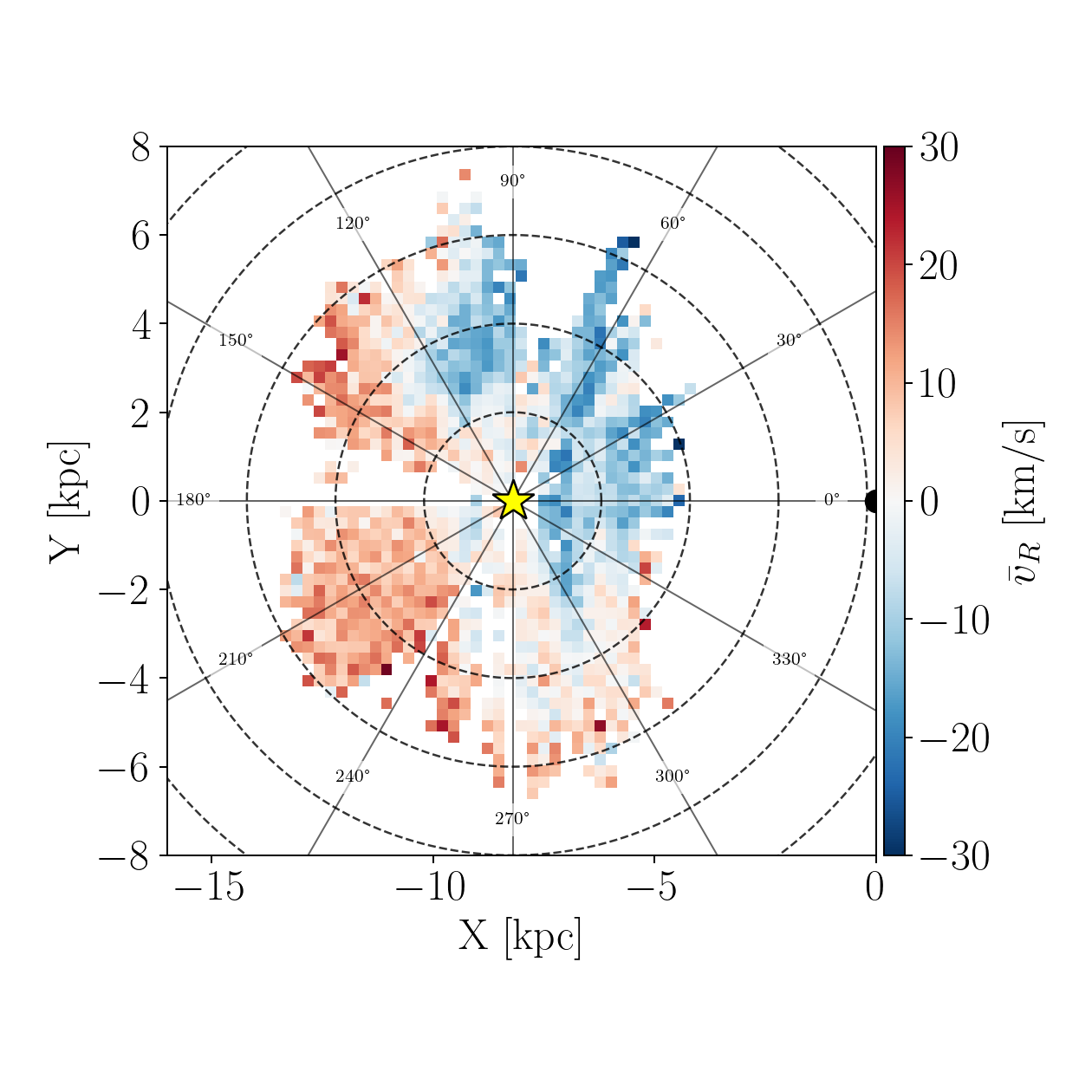}
\caption{Distribution of the Hot Star Sample in the Galactic plane. The left panel shows the density distribution of the sources. The right panel shows the distribution of sources colour-coded by their Galactocentric radial velocity (computed as described in the text). In both panels, the Sun is in $X,Y =$(-8.2,0)~kpc, and it is indicated by the yellow star. The Galactic centre is in $X,Y =$(0,0)~kpc. The dashed circles have radii of 2, 4, 6, 8, and 10 kpc. The solid lines indicate constant Galactic longitudes.}
\label{fig:map-galcenrv}
\end{figure*}

Figure \ref{fig:vphi} shows the azimuthal velocity $v_{\phi}$ of the filtered sample as a function of Galactocentric radius. Within $R \lesssim 5.5$~kpc  we observe hints of  non-axisymmetric dynamics due to the presence of the bar in the Milky Way. The over-densities in the data distribution reflect the selection function of the sample. For instance, the over-densities at $R \sim10$~kpc corresponds to the density enhancement at $X, Y \sim(-10, 2)$ shown in Fig. \ref{fig:map-galcenrv} (left).   Whilst it is not the goal of this paper to fit the rotation curve, we use the fact that stars belonging to the HSS follow cold circular orbits to remove outliers. More precisely, we removed stars whose $v_{\phi}$'s are not between 190 km/s and 250 km/s (orange dashed lines in Fig.  \ref{fig:vphi}). This removes around 5000 objects whose velocities do not closely follow galactic rotation, either due to an erroneous estimate of the radial velocity solution or because they are characterised by peculiar motions deviating from the average azimuthal velocity. These stars in turn could be, for instance, binaries that have not been filtered out from the sample with the RUWE cut, or runaway stars \citep[][]{Hoogerwerf2001}. A detailed analysis of this population is left for future work.

Figure \ref{fig:map-galcenrv} (left) shows the density distribution of the HSS in the Galactic plane. The map is computed by dividing the Galactic plane in spatial bins (spaxels) of constant size of $250 \times 250$~pc. We note that while some of the over-densities in the spatial distribution of the HSS do correspond to young associations and clusters, the overall appearance of the in-plane distribution is driven by the survey selection function. For instance, the lack of sources at negative $X$ values and $Y$ between 0 and 2~kpc is due to lack of observations at $-180 \lesssim l \lesssim -160$ (see Fig.\ref{fig:sky_distribution}). This is partly because the survey is still ongoing, and some regions of the sky have not yet been observed. Conversely, other fields have been observed more frequently due to the survey observing strategy, which includes repeated visits to selected areas; as a result, these fields contain a higher number of observed targets.

We estimated the average Galactocentri radial velocity $\bar{v}_R$ in each spaxel and its associated dispersion $\sigma_R$ by optimising the log-likelihood of the distribution of observed velocities ${v_{R, i}}$ (with uncertainties $\sigma_{v_R, i}$ of the stars located within the j$th$-spaxel). We assumed Gaussian uncertainties, so that the negative log-likelihood to minimise is:
\begin{equation}\label{eq:likelihood}
    \mathcal{L}(\bar{v}_R, \sigma_R)_j = \frac{1}{2}\sum _{i=1}^{N=N_j} \ln(\sigma_R^2 + \sigma_{v_{R,i}}^2) + \frac{(v_{R,i} - \bar{v}_R)^2}{\sigma_R^2 + \sigma_{v_{R,i}}^2}. 
\end{equation}
We considered only the spaxels with $>$5 stars. 
Figure  \ref{fig:map-galcenrv} (right) shows the distribution of the HSS colour-coded by $\bar{v}_R$. Figure \ref{fig:vR_error} shows the velocity dispersion $\sigma_R$ of the sample. In Fig. \ref{fig:map-galcenrv}, stars moving on average toward the Galactic centre are coloured in blue, whereas outwards moving stars are coloured in red.  Radial motions within the solar circle (between the Sun and the Galactic centre) are predominantly directed inwards (negative Galactocentric radial velocities). Radial motions outside the Solar circle ($R \gtrsim 8$~kpc) are instead predominantly directed outwards (positive Galactocentric radial velocities).
The inward motions  show strong asymmetric signatures. These features extend over several kpc, with the most prominent drawing an arc-like segment extending almost over the entire reach of the sample. We further discuss this result in Section \ref{sec:discussion}.

\subsection{Ages}\label{sec:ages}
We estimated the ages of the HSS sample members following the methodology proposed by \citet[][JL05]{Jorgensen2005} and more recently \cite{howes2019} and \cite{Sahlholdt2021}. Using the same notation as JL05, we write  the posterior probability $f(\tau, Z, m)$ for the log-age $\tau$, metallicity $Z$,  and mass $m$, as:
\begin{equation}
    f(\tau, Z, m) \propto f_0(\tau,Z, m)\mathcal{L}(\tau, Z,m),
\end{equation}
where  $f_0(\tau,Z, m)$ is the prior on log-age, mass and metallicity, and $\mathcal{L}(t, Z,m)$ is the likelihood function. Integrating with respect to $m$ and $Z$ gives $f(\tau)$,
the posterior pdf of $\tau$, which  summarizes
the available information concerning the log-age of the star.

We wrote the prior $f_0(\tau,Z, m)$ as the product of three independent components:
\begin{equation}
    f_0(\tau,Z, m) = \psi(\tau) \xi(m) \phi(Z). 
\end{equation}

We chose a uniform prior in age $t$, $\psi(t)=c$, therefore as probability densities need to be conserved:
\begin{align}
    \psi(\tau)~d\tau &= \psi(t)~dt = c~dt  \nonumber \\
    \psi(\tau) &= c \frac{dt}{d\tau} = \tilde{c}10^{\tau}, 
\end{align}
with $\tau = \log_{10}(t)$. For simplicity we set $\tilde{c} = 1$.

The prior on the metallicity $Z$, or more precisely on the iron abundance [Fe/H], is given by a  Gaussian distribution centred around the mean metallicity at a given Galactocentric radius:
\begin{equation}
    \phi(Z) = \frac{1}{\sqrt{2\pi} \sigma } \exp  \frac{(Z - Z_{\mathrm{exp}})^2}{2\sigma_{Z}^2} .
\end{equation}
We determined the mean $Z_{\mathrm{exp}} = Z_{\mathrm{exp}} (R_{\mathrm{Gal}})$ by using the relation found by \citet{Hackshaw2024}:
\begin{equation}
    Z_{\mathrm{exp}} = -0.0678~R_{\mathrm{Gal}} +0.546,
\end{equation}
and we assumed a constant $\sigma_Z = 0.1$~dex. 

Finally, we assume a simple power-law for the stellar mass prior:
\begin{equation}
    \xi(m) \propto m^{-\alpha},
\end{equation}
where $\alpha = -2.7$, which is representative for the empirical IMF at $m > 1 M_{\odot}$ \citep[][]{Kroupa2002}.

For each star, we adopt a set of observables — effective temperature ($T_{\mathrm{eff}}$), surface gravity ($\log g$), and absolute magnitude in the $K$-band ($M_K$) — and compute the likelihood assuming Gaussian uncertainties on these parameters:
\begin{equation}
    \mathcal{L}(\mathbf{d} | \tau, Z, m) \propto \prod_i \exp  \frac{(d_i - q_i(\tau, Z, m))^2}{2\sigma_i^2} ,
\end{equation}
where $\mathbf{d} = ({T_{\mathrm{eff}}, \log g, M_K})$ are the observed values with uncertainties $\sigma_i$, and $q_i(\tau, Z, m)$ are the corresponding model predictions for age $\tau$, mass $m$,  and metallicity $Z$, interpolated from stellar isochrones.

Integrating over $m$ and $Z$, the posterior probability of log-age $\tau$ can be written as:
\begin{equation}
    f(\tau) \propto \psi(\tau)\mathcal{G(\tau)},
\end{equation}
where:
\begin{equation}
    \mathcal{G}(\tau) = \int\int \mathcal{L}(\mathbf{d}|\tau, Z, m)~\xi(m)~\phi(Z)~dm  dZ.
\end{equation}
In practice, we approximated the integral over mass $m$ and $Z$ as:
\begin{equation}
    \mathcal{G}(\tau_k) \propto \sum_{i,j} p(Z_j)~p(m_{ijk})~\mathcal{L}(\tau_k, Z_j, m_{ijk})~dm_{ijk} ~dZ,
\end{equation}
where  $m_{ijk}$ are the initial-mass values along each isochrone ($\tau_k$, $Z_j$). Finally, we multiplied the marginalized likelihood by the prior on log-age $\psi(\tau)$ and normalized the posterior distribution. We evaluated $f(\tau)$ by using MIST \citep[][MIST version 1.2, with Initial v/v\_crit = 0.4]{Choi2016, Dotter2016} isochrones equally spaced (0.05 dex) in log-age, from 1 Myr to 1 Gyr, and with [Fe/H] from -0.5 to +0.4 dex, in steps of 0.1 dex. We use the 50th percentile of $f(\tau)$ as our age estimator, and the semi-difference between the 84th and 16th percentile of $f(\tau)$ as our uncertainty estimator.

\begin{figure}
    \centering
    \includegraphics[width=\linewidth]{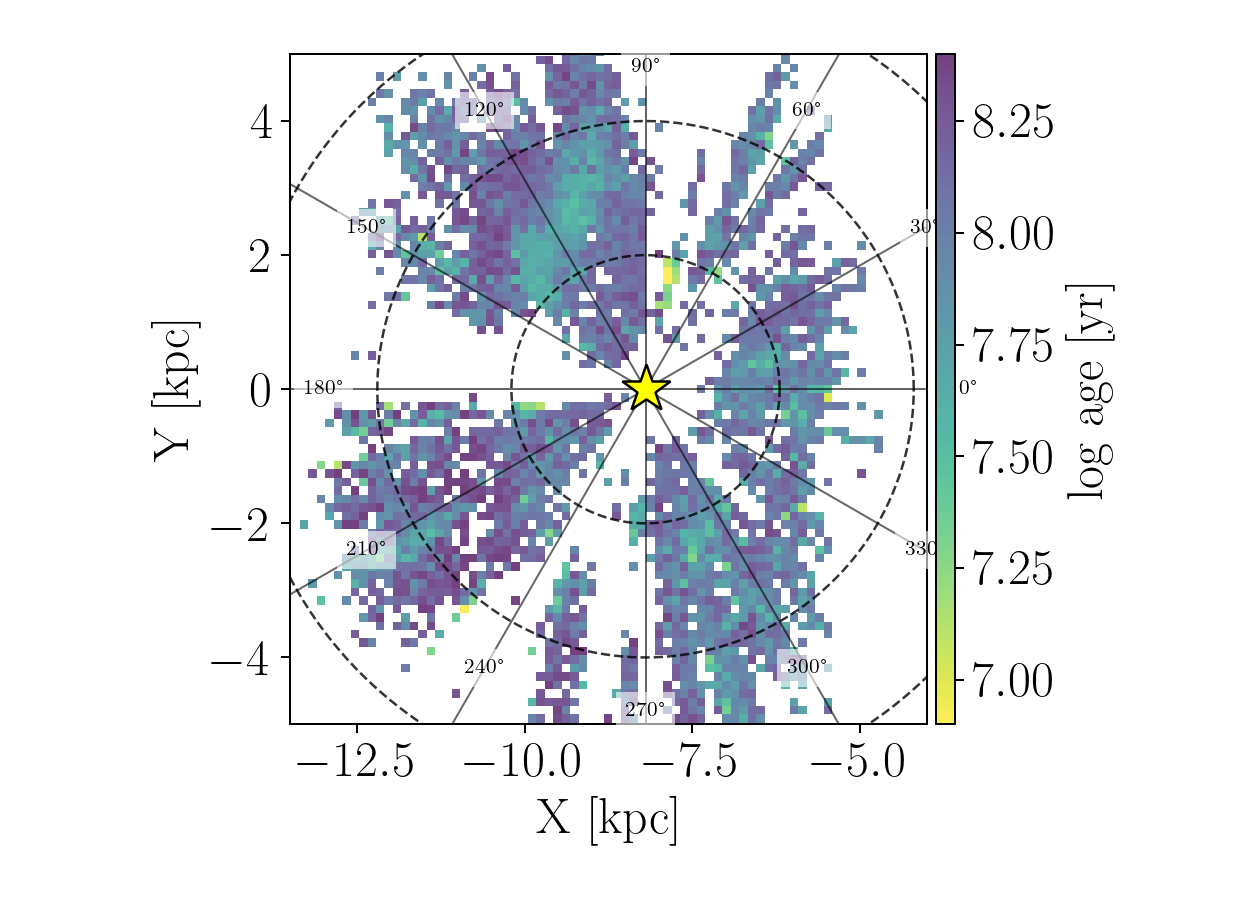}
    \caption{Mean log-age map  of the Hot Star Sample in the Galactic plane. The mean ages are computed only for spaxels with more than 5 stars. The dashed circles have radii of 2, 4, and 6~kpc. The yellow star indicates the Sun's position. The solid lines indicate constant Galactic longitudes.}
    \label{fig:age_plane_all}
\end{figure}

Figure \ref{fig:age_plane_all} shows a map of mean ages in the Galactic plane, in spaxels of size 125x125 pc. Only spaxels with more than 5 stars are shown. Wide clumps of stars with young ages (log-age $\lesssim$ 7.5 $\sim$ 30 Myr) are visible. These correspond to star formation regions and will be further discussed in Section \ref{sec:discussion}. The 85\% of the HSS is younger than $\sim$320 Myr (log-age = 8.5 dex). The median age uncertainty of the entire sample is 15~Myr. For stars younger than 30~Myr (log-age = 7.5), the median age uncertainty is $\sim$4 Myr, for stars older than $\sim 150 $~Myr (log-age = 8.2) is $\sim30$~Myr.

\begin{figure*}
    \centering
    \includegraphics[width=0.49\linewidth]{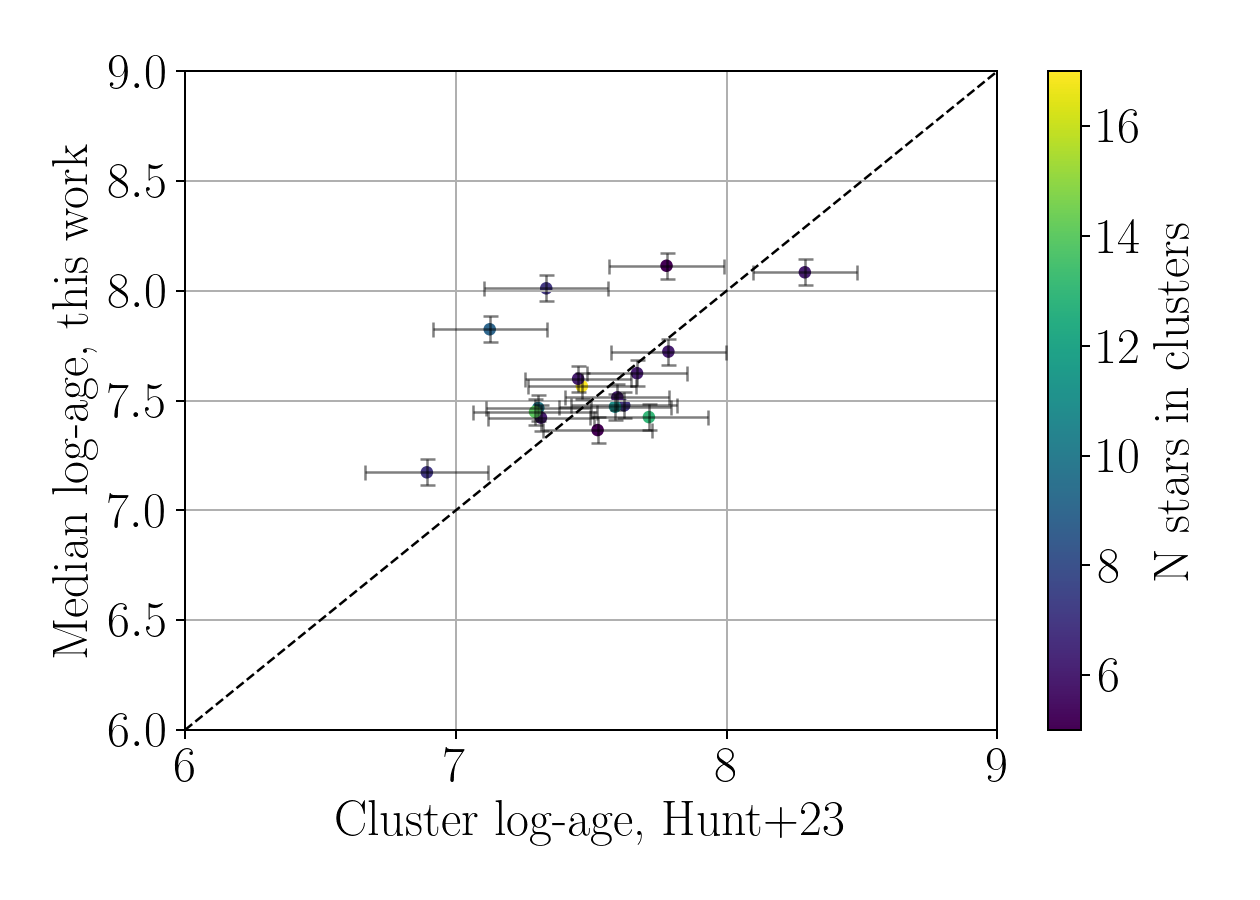}
    \includegraphics[width=0.49\linewidth]{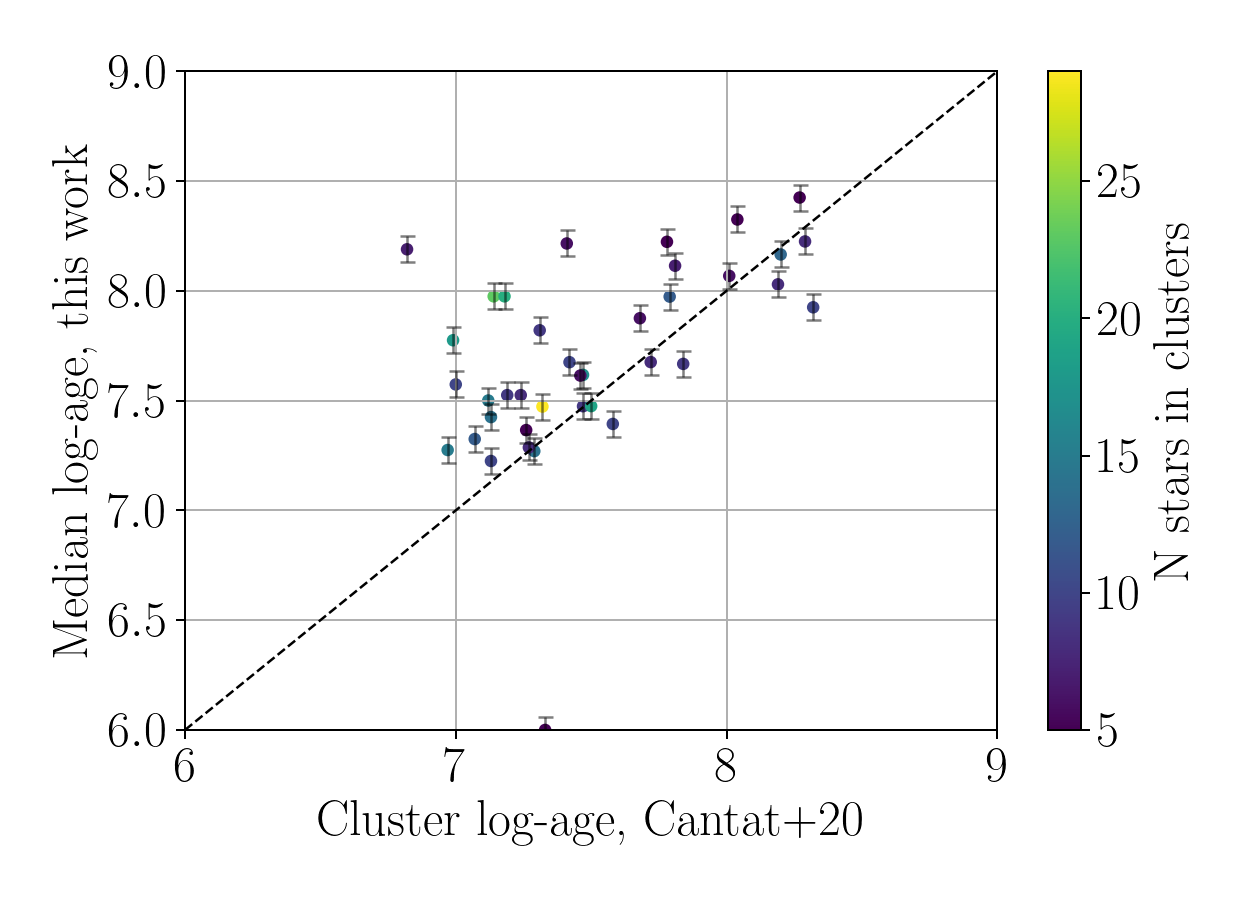}
       \caption{Comparison between the ages for clusters determined in this work and by \cite{Hunt2023} and \cite{Cantat2020}. The colour bar represents the number of hot stars in the cluster. The dashed line represent the 1:1 relation in both panels. }
    \label{fig:compare_age}
\end{figure*}

We validated our age estimates by comparing with the age estimates for clusters provided by \citet[][H23]{Hunt2023} and \citet[][CG20]{Cantat2020} . We cross-matched our HSS with the cluster members of both H23 and CG20 (with membership probability > 50\% and 90\% respectively), we evaluated the posterior distributions $f_{\star}(\tau)$ of the "Hot Stars" (HS) cluster members and multiplied them together (under the reasonable assumption that cluster members are coeval) and computed the HS cluster age as the 50th percentile of the total posterior distribution $f_{\mathrm{HS}} = \prod_{\star} f_{\star}(\tau)$. We used only clusters with at least five HS with posteriors  narrower than 0.3 dex in log-age  (which is around a factor of 2 on age).
Figure \ref{fig:compare_age} shows the comparison between the HS cluster ages and H23 (left) and CG20 (right). The comparison with H23 and CG20 is satisfactory. With respect to CG20, our ages seem to be slightly over-estimated (although CG20 does not provide errors on their age estimates). With respect to H23 they agree well, except for a few outliers. The uncertainties of age estimates are much smaller than those estimated by H23: this might mean that our method tends to underestimate uncertainties. In any case the ages determined in this work are intended as youth indicators  rather than exact age determinations.

\section{Discussion}\label{sec:discussion}
Figure~\ref{fig:map-galcenrv} (right panel) shows that massive young stars in the Milky Way disc have remarkably strong radial motions, coherent over several kpc. These large-scale kinematic patterns in the $X-Y$ plane could be a signature of the dynamical response of young massive stars to the Milky Way’s spiral structure, but could also arise from other phenomena, such as the influence of the bar, resonant interaction, or incomplete phase mixing following some perturbation. In this Section we put our results in the context of several observational studies and simulations, and compare the motions of younger and older sources. 

\subsection{Comparison with spiral arm tracers}
\begin{figure*}
    \centering
    \includegraphics[width=0.33\textwidth]{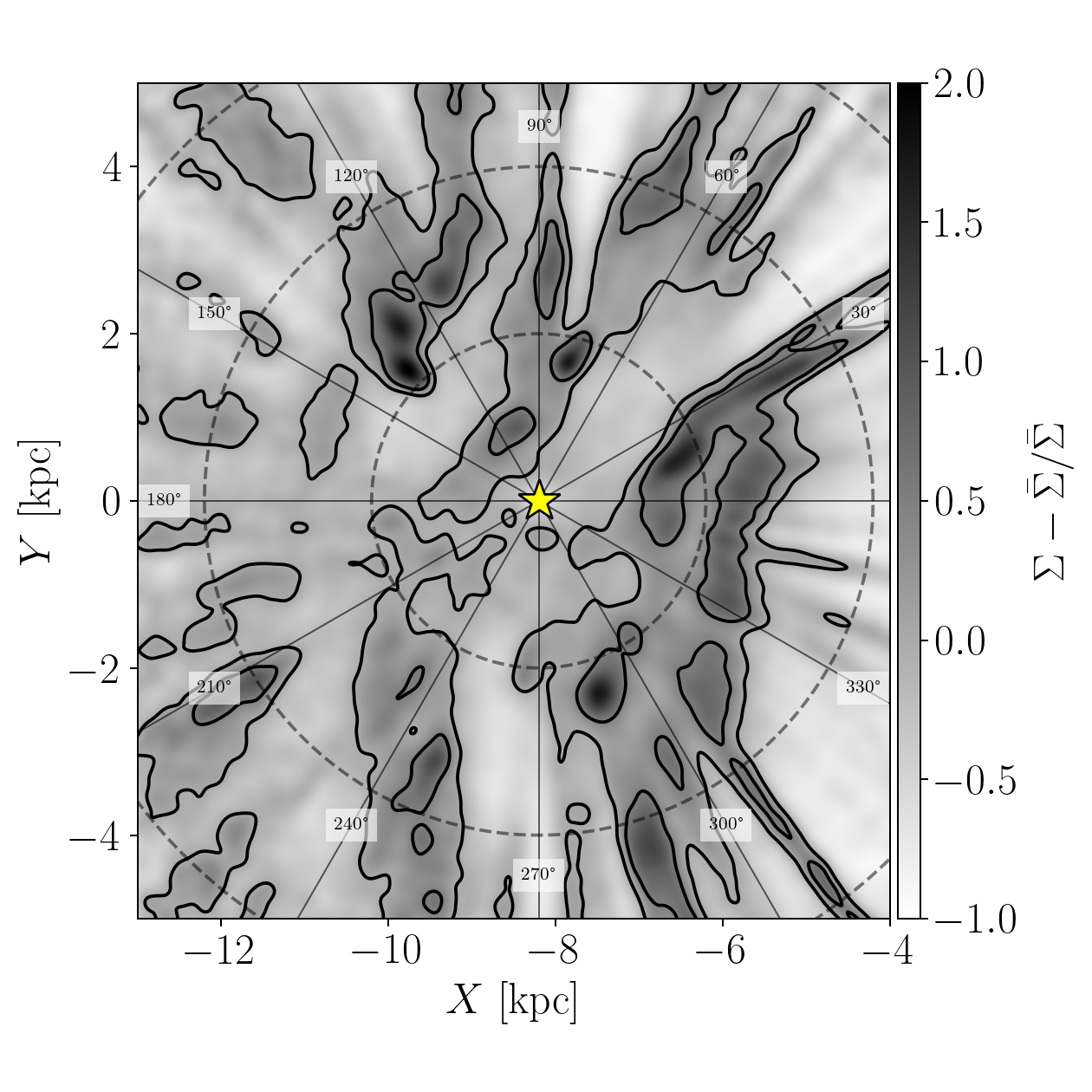}
    \includegraphics[width =0.33\textwidth]{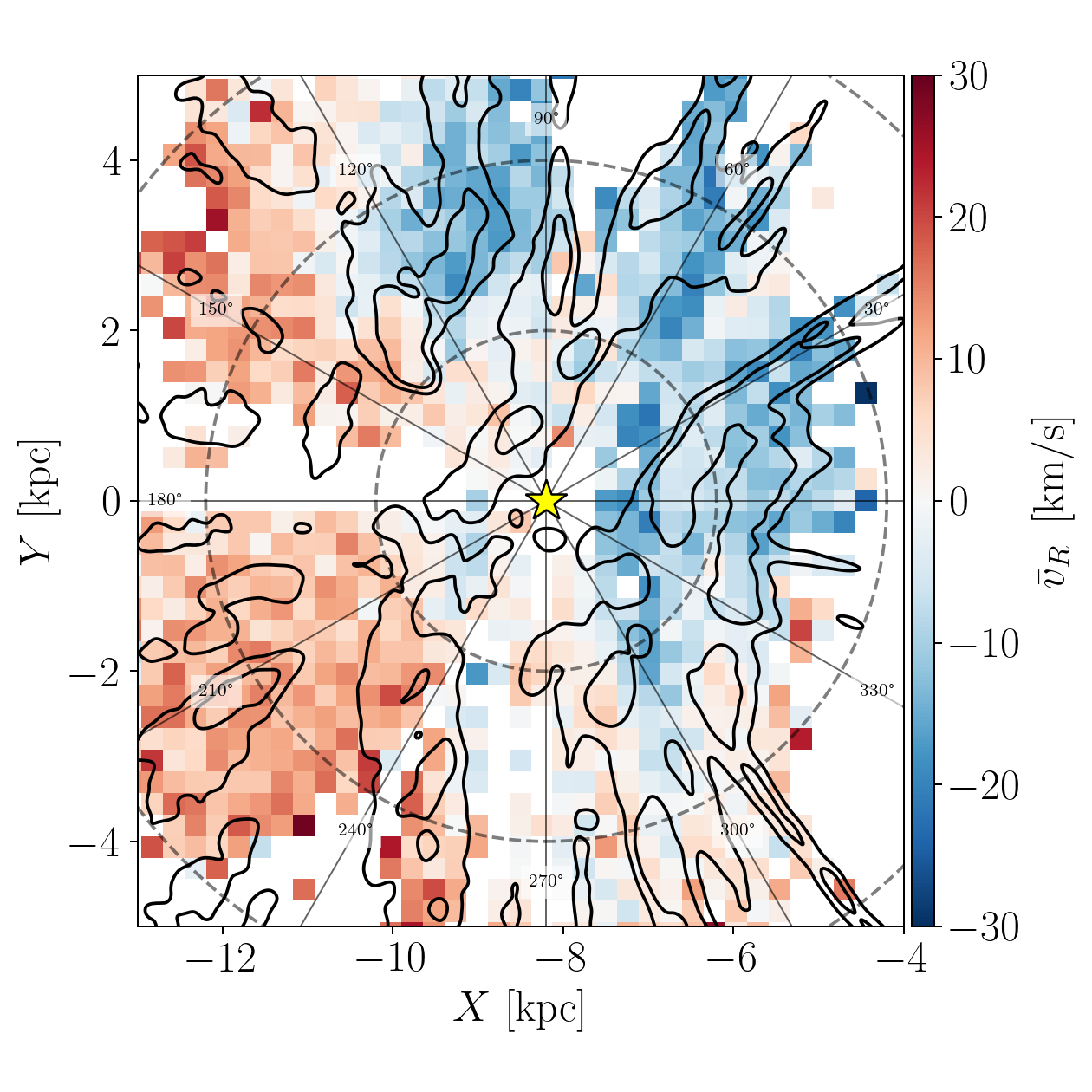}
    \includegraphics[width =0.33\textwidth]{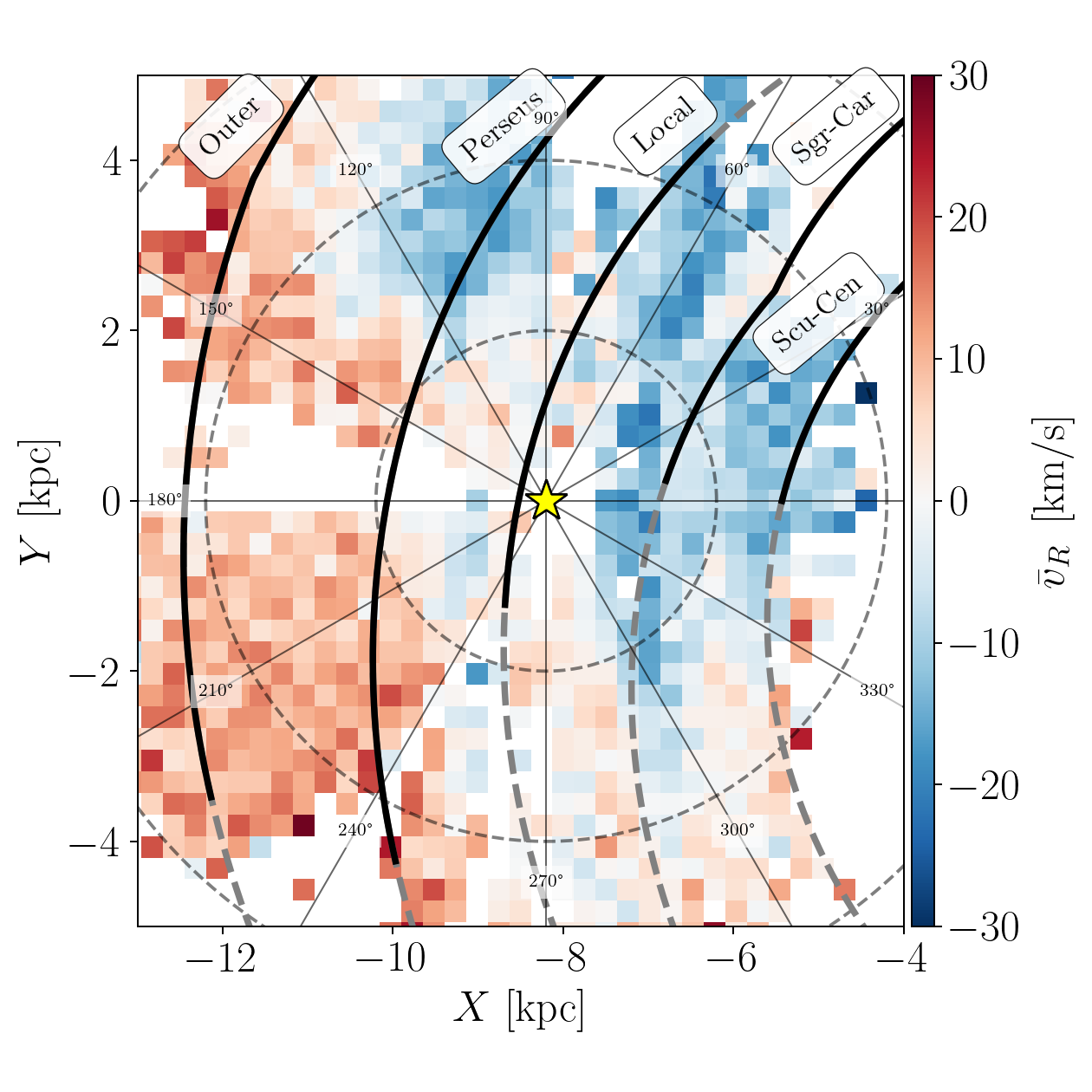}
    \caption{Comparison between the density distribution of OB stars, the spiral arm model from  \cite{Reid2019}, and the radial velocity asymmetries.  In the left panel and central panels, the solid black contours  represent the  0, and 0.5 density contrast levels. In the right panel, the spiral arms are named following \cite{Reid2019}: Outer arm; Perseus arm; Local arm;  Sagittarius–Carina arm; Scutum–Centaurus arm. In the three panels, the yellow star indicates the Sun's position in $X,Y = -8.2,0$~kpc, and the Galactic centre is located in $X,Y = 0,0$~kpc. The dashed lines represent circle radii of  2, 4, and 6 kpc.  The solid lines indicate constant Galactic longitudes.}
    \label{fig:compare_rv}
\end{figure*}

Figure \ref{fig:compare_rv} shows a comparison between the $\bar{v}_R$ map and the density distribution of the OB star sample of \cite{Zari2021} and the spiral arm models derived by \cite{Reid2019}. The left panel of Fig. \ref{fig:compare_rv} shows the density contrast of the OB sample, defined as:
\begin{equation}\label{eq:deltaSigma}
\Delta_{\Sigma} = \frac{\Sigma - \bar{\Sigma}}{\bar{\Sigma}}.
\end{equation}
We computed $\Delta_{\Sigma}$ by using a similar procedure as \cite{Poggio2021}. We derived the 3D density $\rho(X, Y, Z)$ of the OB sample by smoothing the spatial distribution of stars with a 3D Gaussian filter with  bandwidth  $w = 10$ pc, and an average density $\bar{\rho}(X,Y,Z)$ by smoothing with a larger bandwidth $w = 100$~pc. We truncated both filters at 3$\sigma$. We projected the distribution on the Galactic plane by
integrating over $Z$, to derive $\Sigma = \Sigma(X,Y)$ and $\bar{\Sigma}(X,Y)$, and finally computed  $\Delta_{\Sigma}$. The OB star over-densities have been interpreted as spiral arm segments. Surprisingly, the middle panel of Fig. \ref{fig:compare_rv} shows that the kinematic structures are not aligned with the OB star density. The black lines in right panel of Fig. \ref{fig:compare_rv} show the spiral arm model derived by \cite{Reid2019}. While the alignment is not perfect, the model seem to follow the kinematic structure slightly more closely. 

\begin{figure*}
\includegraphics[width = 0.49\textwidth]{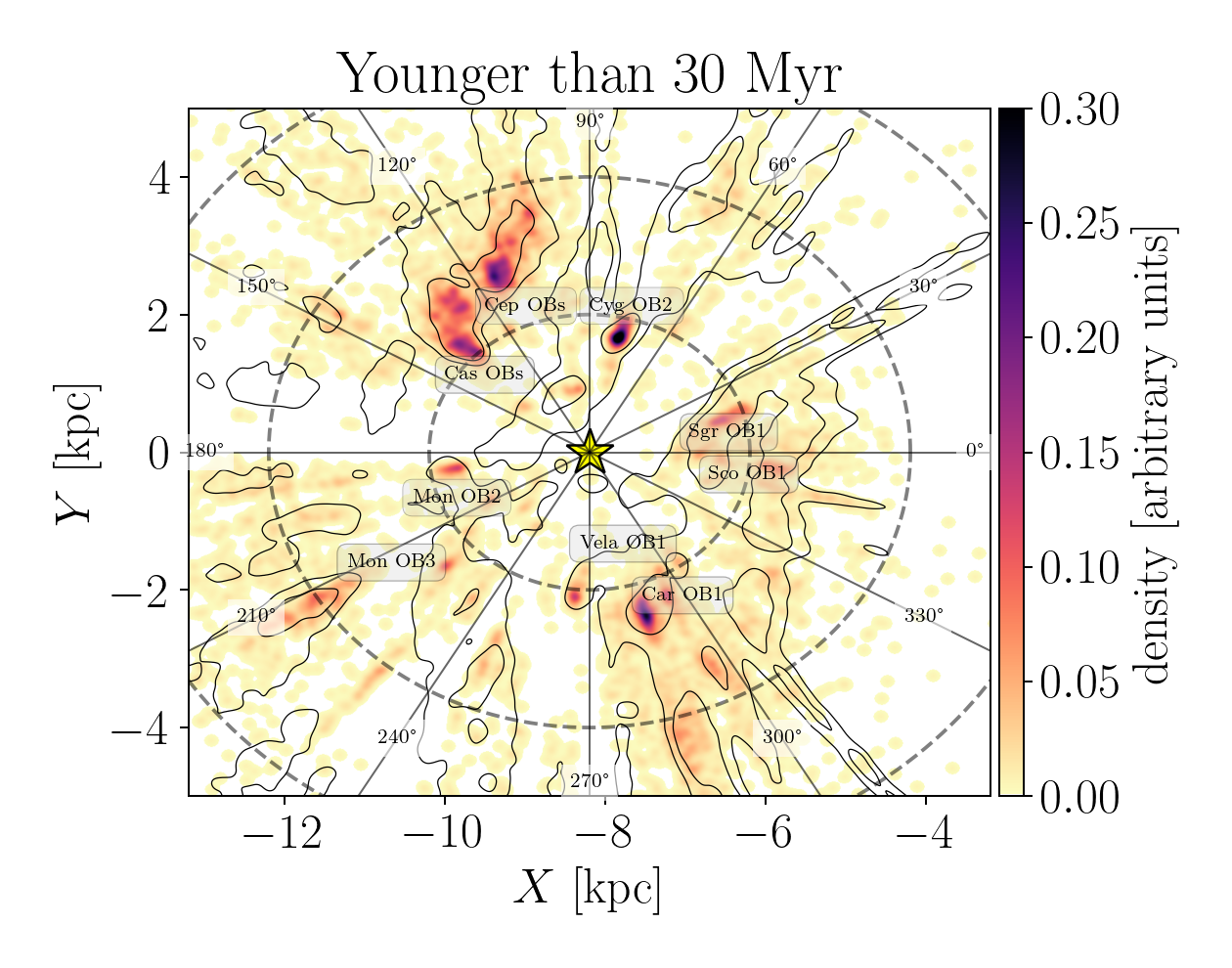}
\includegraphics[width = 0.49\textwidth]{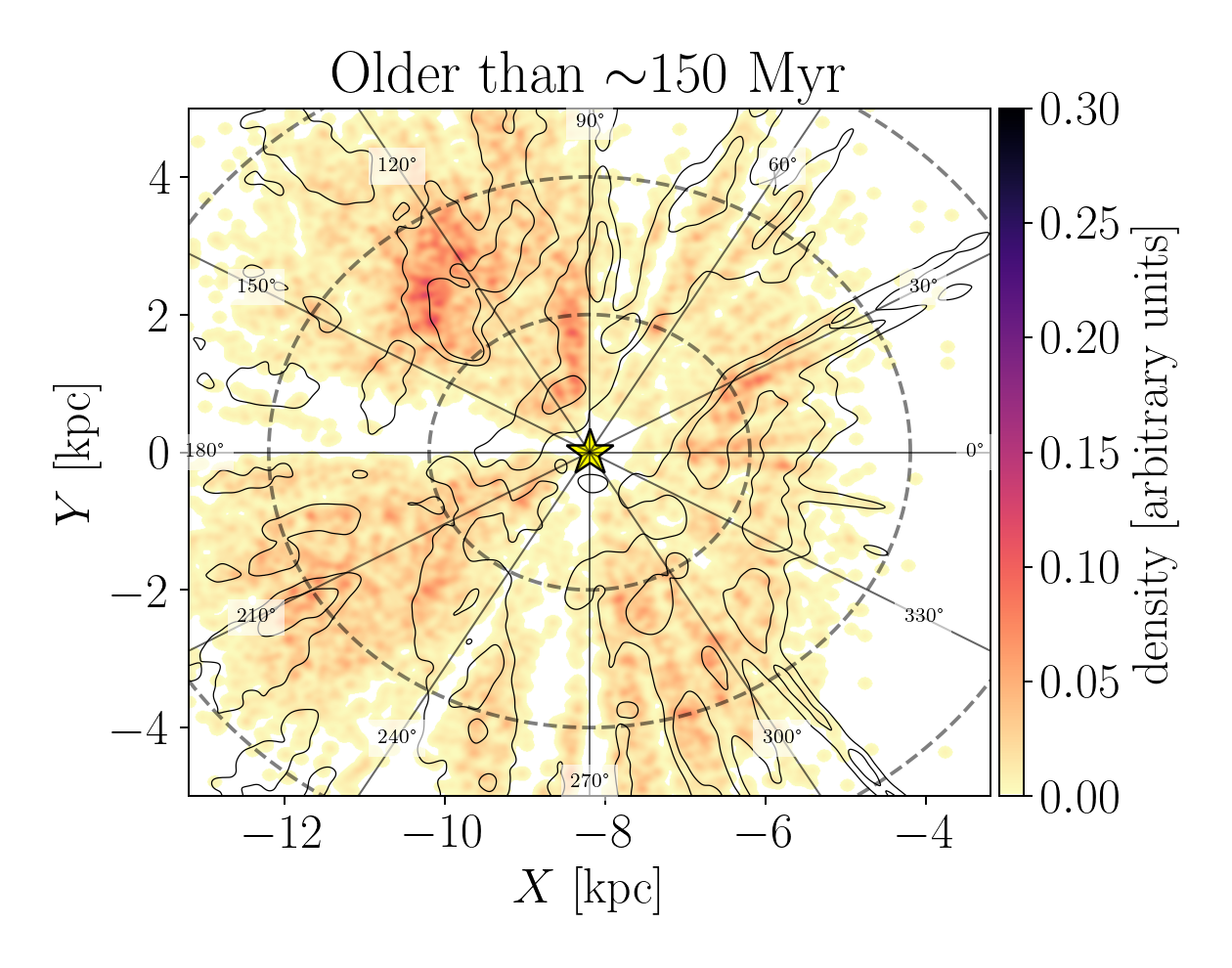}
\caption{Density distribution of young ($<$ 30 Myr), left and older ($>$ 150 Myr) sources. In the left panel, star forming regions are indicated. The solid contour lines represent the density distribution of all sources (Fig. \ref{fig:compare_rv}, left), and are computed as described in Sec. \ref{sec:discussion}. The labels highlight some of the more prominent OB associations \citep[their locations are from Table 1 of][]{Wright2020}.  The solid lines indicate constant Galactic longitudes.}
\label{fig:plane_ages}
\end{figure*}

Figure \ref{fig:plane_ages} shows the density distribution of young ($\lesssim$ ~30~Myr, left, 7,704 stars) and older ($\gtrsim 150$~Myr, right, 12,169 stars) stars in our sample. The solid lines correspond to the $\Delta \Sigma$ contours computed with Eq. \ref{eq:deltaSigma} and showed in Fig. \ref{fig:compare_rv}. The young star distribution is clustered, and shows several clumps that can be associated to OB associations and star formation complexes, such as those towards Cygnus \citep{Wright2016_Cygnus, Berlanas2020_Cygnus}, Cassiopeia \citep[W3,W4,W5,][]{Gouliermis2014_Cassio}, Carina \citep[Tr14, Tr15, Tr16,][]{Berlanas2023_Carina}, and Sagittarius and Serpens \citep[M20, M17,][]{Povich2007_M17, Kuhn2021}. The density distribution of the youngest sources  traces  the density enhancements of the  overall OB star distribution \citep[consistent with Fig. 4 of ][]{Zari2023}. The older stars (> 150 Myr) show less prominent density variations. The residual over-densities might be artifacts, due to  uncertainties in the age estimates and to the survey selection function (that is, in some fields more targets are observed than in others), or might still reflect spiral structure \citep[see][]{Zari2023}.

While a connection between spiral arms and the kinematics of nearby stars and gas is expected, establishing a direct link is not straightforward. 
The youngest stars, formed within spiral arms, are expected to inherit the motions of their parent molecular clouds, which typically rotate with relatively low random velocities. The bulk motion of star-forming gas within spiral arms is influenced by shear arising from gravitational torques. In the reference frame of the spiral pattern, gas located on the trailing side of the arm moves outward, while gas on the leading side moves inward~\citep[see, e.g.][]{2014MNRAS.443.2757K}. If shocks are present along the leading edge of the spiral arms, raising gas pressure, this may lead to enhanced star formation in these regions~\citep{2013ApJ...779...45M}. Consequently, stars formed on the leading side are expected to be more numerous and to exhibit a net negative radial velocity. 
The amplitude of the radial velocity component of newly formed stars depends on both the spiral pattern speed and the pitch angle. Therefore, the observed mean negative radial velocities inside Galactocentric radii of <9~kpc (as shown in Figs.~\ref{fig:map-galcenrv} and~\ref{fig:compare_rv}) suggests that a significant fraction of these stars formed in spiral arms and still retain the bulk motion of their natal gas. This idea is also supported by recent studies of young stellar clusters \citep{Swiggum2025}, which showed that clusters formed near spiral arms exhibit systematic radial motions that might track the passage of the arm and the associated gas compression.

The observed spatial offset between spiral arms traced by maser sources \citep[][]{Reid2014, Reid2019} and those traced by young stars can be attributed to differences in the orbital frequencies of stars formed at various locations along the spiral arms, relative to the pattern's corotation. In the case of rigidly rotating spiral patterns, this results in the youngest stars being located within the spiral arms, while somewhat older stars ($\sim 100$~Myr) appear downstream in the inter-arm regions~\citep{2010MNRAS.409..396D}. While this picture is intuitively appealing, it breaks down for tidally induced or flocculent spiral structures, where a clear age gradient is not expected. Although one could, in principle, model the displacement of stars as a function of their age, such an approach requires assumptions about the historical structure, kinematics and the origin of the Milky Way spiral arms, an analysis that lies beyond the scope of the present study.

\subsection{Comparison with Aurigaia}
\begin{figure*}
    \includegraphics[width = 0.48\hsize]{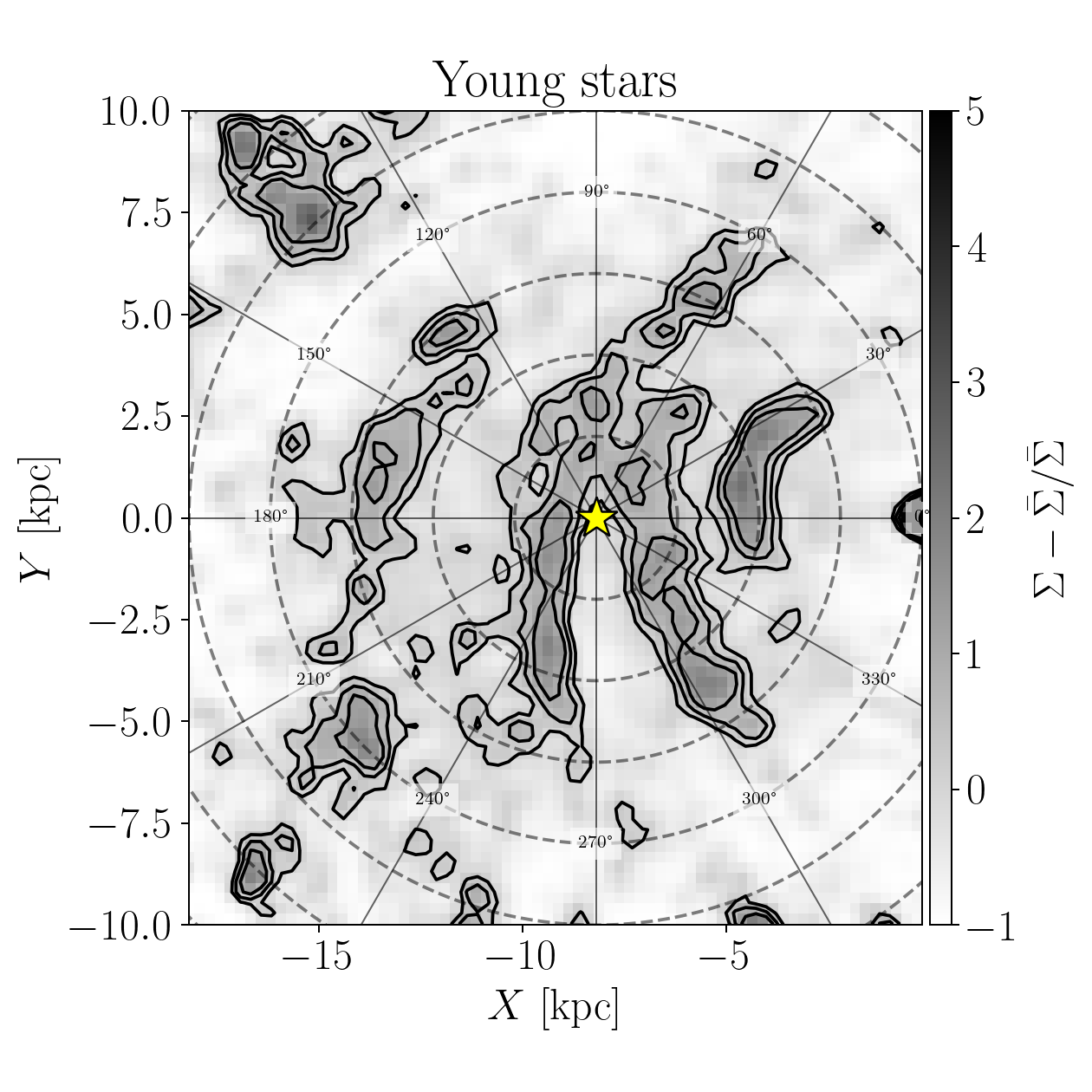}
     \includegraphics[width = 0.48\hsize]{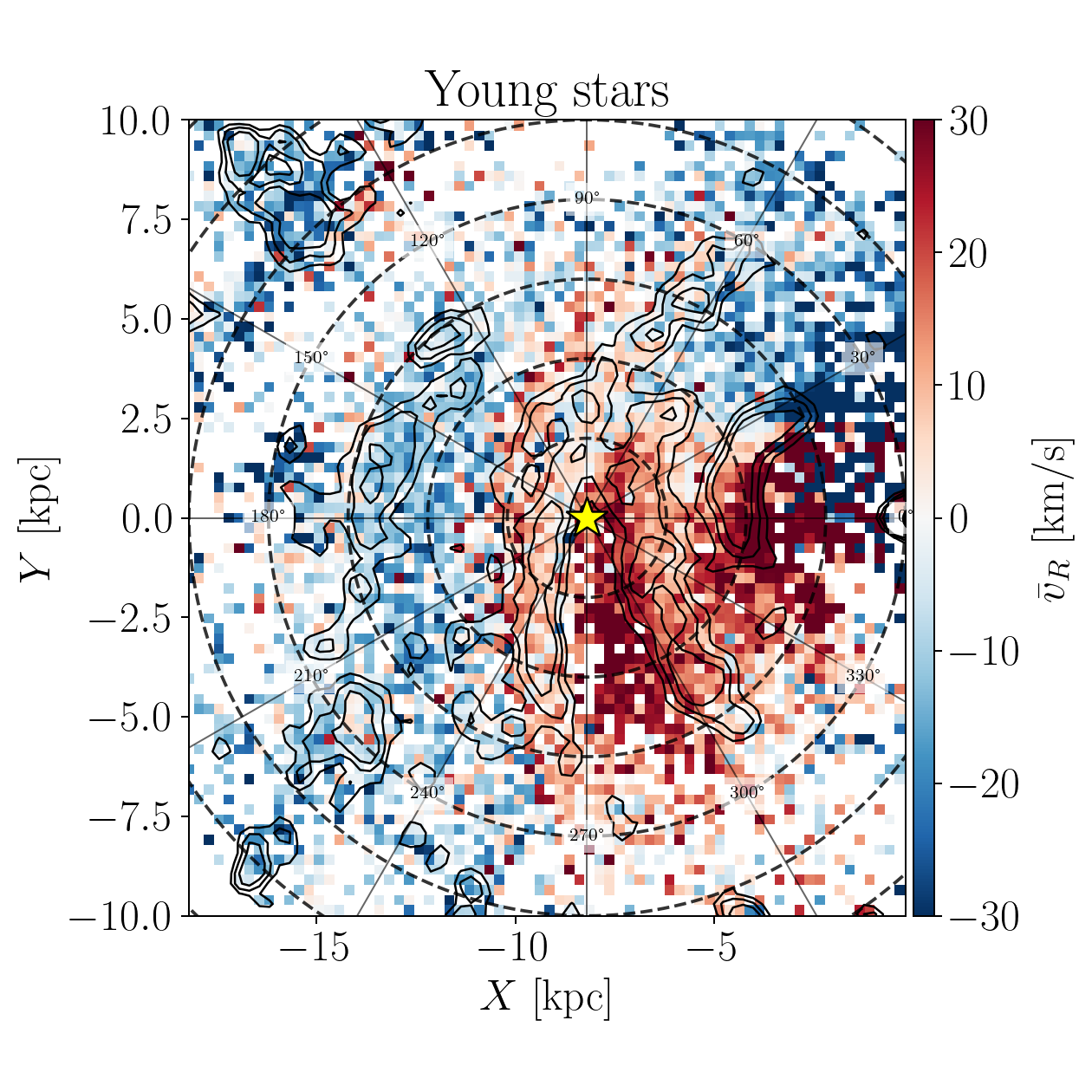}\\
    \includegraphics[width = 0.48\hsize]{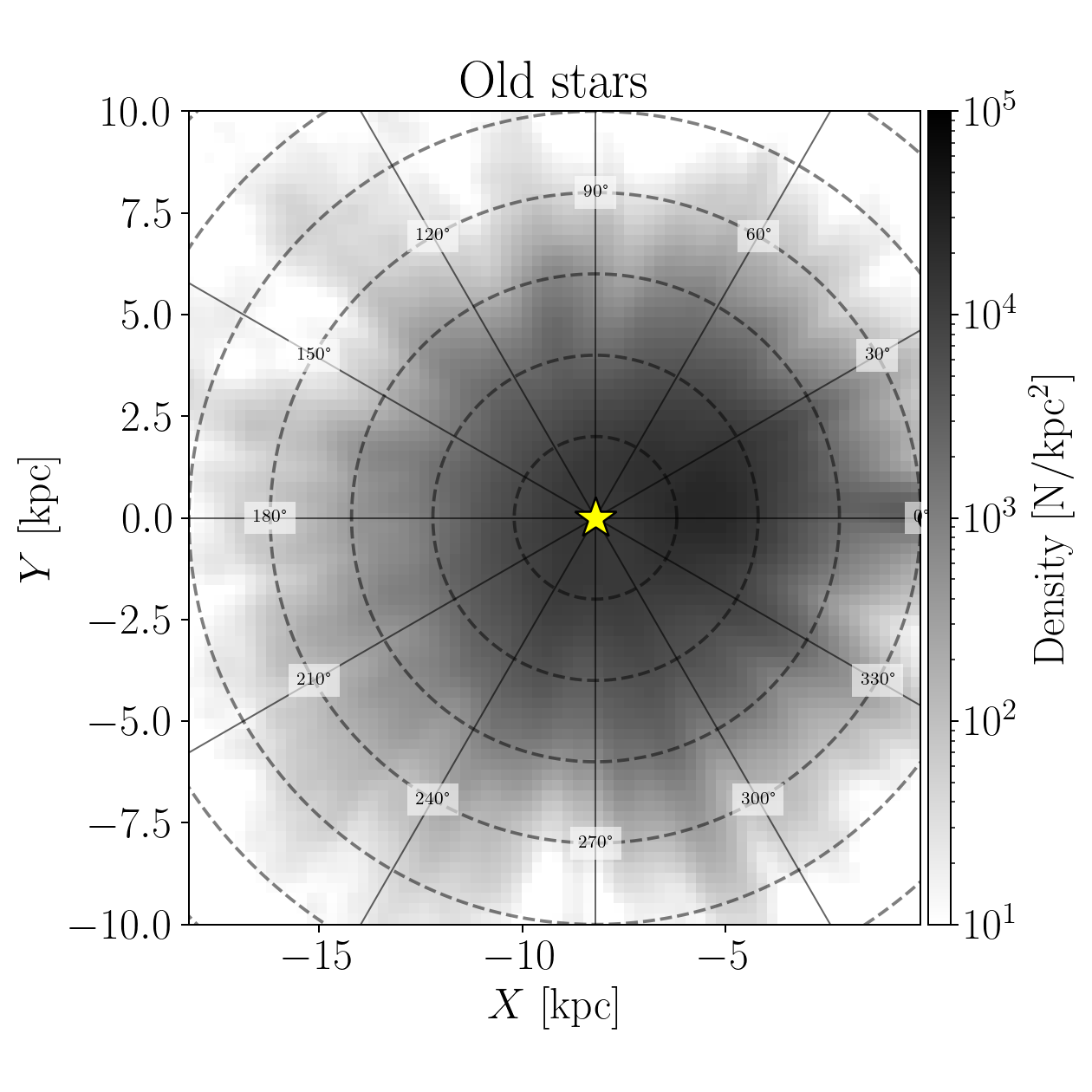}
     \includegraphics[width = 0.48\hsize]{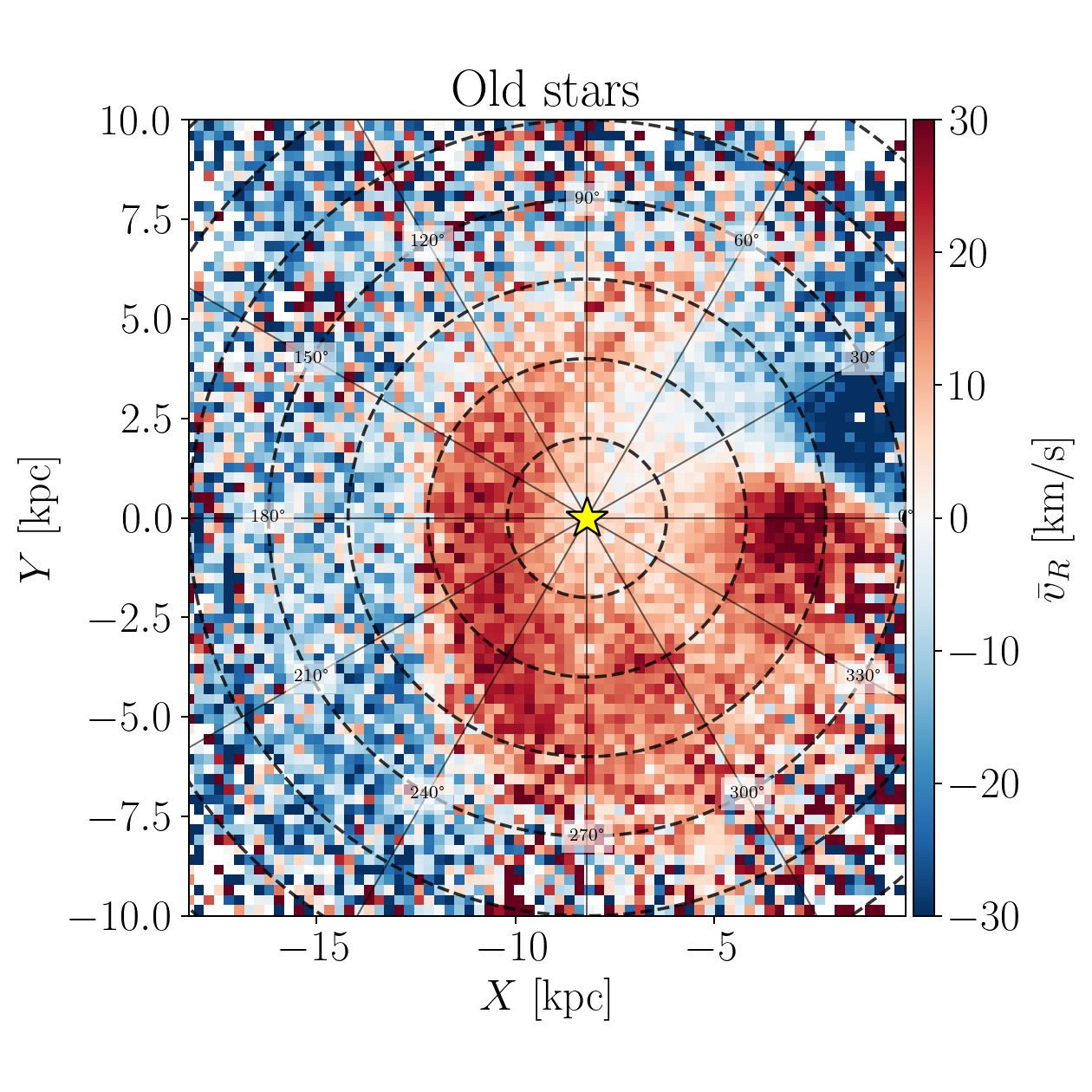}
   
\caption{Density distribution (left) and Galactocentric radial velocities (right) of the \texttt{gaiauriga} mock stellar catalogue for young (top) and old (bottom) sources. The black lines outline  the contours of the density distribution. The black dot indicates the Galactic centre, the yellow star the Sun's position.  The dashed black lines mark concentric circles at $\Delta R=2$ kpc. The $\bar{v}_R>$ feature for  $R \lesssim 5$~kpc in the right panels  is due to the effect of the bar on the kinematics of sources in the inner galaxy.  The solid lines indicate constant Galactic longitudes.}
\label{fig:auriga}
\end{figure*}

To gain better insights on our results, we compared our observational data with the mock stellar catalogues generated from the \texttt{AURIGA} cosmological simulations\footnote{at this link: \url{https://wwwmpa.mpa-garching.mpg.de/auriga/gaiamock.html}} \citep[][]{Grand2018}.  Here we show the HITS mocks catalogue generated using
the SNAPDRAGONS code by \cite{Hunt2015} with the following
parameters: Halo 6, resolution level 3, with extinction, but we found analogous results by using all the available HITS mock catalogues (Halo 16, 21, 23, 24, and 27).
The angle between the Sun and the Galactic Centre line and the major axis of the bar is 30 deg. We used the following ADQL script to replicate the HSS selection function:  
\begin{align}
& \texttt{Gmagnitude} < 16.0 ~\mathrm{mag} \nonumber \\
& \texttt{Age} < 0.5 ~\mathrm{Gyr}\nonumber \\ 
& \texttt{EffectiveTemperature} > 10,000 ~\mathrm{K}\nonumber
\end{align}
Figure \ref{fig:auriga} (top left) shows the density distribution of the sources we selected. The \texttt{aurigaia} catalogue is limited to $V < 16$ mag, therefore the coverage of the Galactic plane is very similar as our observed sample. The density distribution traces the spiral structure of the simulated galaxy, with visible over-densities corresponding to segments of spiral arms.  Figure \ref{fig:auriga} (top right) shows the same distribution, colour-coded by  Galactocentric radial velocities.  The simulation shows radial velocities asymmetries, with comparable amplitude as those shown in Fig. \ref{fig:map-galcenrv}. The bi-symmetric feature on either side of the Galactic Centre, with negative and positive values on each side of the apparent major axis of the bar, is generated by the Galactic bar.
The qualitative agreement  between our data and the \texttt{aurigaia} simulation suggests that radial velocities asymmetries are  a common feature in spiral galaxies. There are however a few differences. In the simulation, the stellar over-densities within $\sim6$~kpc from the Sun have coherent outwards motions, while the stellar overdensity at $R\sim 6-8$~kpc towards the outer Galaxy exhibits mostly inwards motions.  In our observations instead, stars seem to be moving mostly towards the Galactic centre, and only a few stellar overdensities move outwards (such as those at $X < -10, Y < 0$~kpc. These discrepancies might depend on the detailed properties of spiral arms in the simulations (position, pattern speed,...). Even in the simulation, the spiral arm segments do not all move at constant speed, with  over-densities moving with both positive and negative velocities.

\subsection{Comparison with kinematic studies}\label{sec:comparison_kin}
We now compare our results with two observational studies based on young stars: D23 and \citet[][P24]{Poggio2024}. 
D23 found that the velocity field of the OB stars shows streaming motions that have a characteristic length similar to the spiral arm density. The region sampled by the OB sample in D23 extends up to distances of $\sim$2 kpc, thus the streaming motions in the OB stars map mainly the Local Arm. Conversely, the sample that we used in this study is highly incomplete at distances below 2 kpc. This is due both for the selection function of the sample (see Z21), and the observing strategy \citep{Medan2025}. However, our sample extends even beyond 6 kpc towards certain lines of sight, allowing us to map the velocities of Hot Stars on a much larger volume than D23. Thus, comparing the two samples in the same volume is - at the moment - almost impossible. As a sanity check, we verified that the stars in common between the two samples have comparable $v_R$ (see also Section \ref{sec:data}).
P24 studied the median Galactocentric radial velocity for a sample of young giants (age $\lesssim$ 100 Myr) and Cepheids and found a strong feature in the $\bar{v}_R$ velocity field. They interpreted this feature as the signature of a wave propagating towards the outer disc, and suggested that the kinematics of young stars in the outer disc is dominated by this outwards motion. Figure \ref{fig:map-galcenrv} shows that towards the outer disc, at distances larger than $\sim$2~kpc from the Sun, the radial motions of the HSS are directed outwards, similar to P24,  albeit with stronger $\bar{v}_R$ variations ($-30 < \bar{v}_R < 30$ km/s as compared to $-15 < \bar{v}_R < 15$ km/s found by P24). Towards the inner disc, we detect stronger inwards motions than reported by P24. This is likely due to the differences in the samples used in the two studies. For instance, the spatial distribution of the young giant sample in Fig. 4 of P24 presents different features than the one showed in Fig. \ref{fig:map-galcenrv}.

\begin{figure}
    \centering
    \includegraphics[width=\hsize]{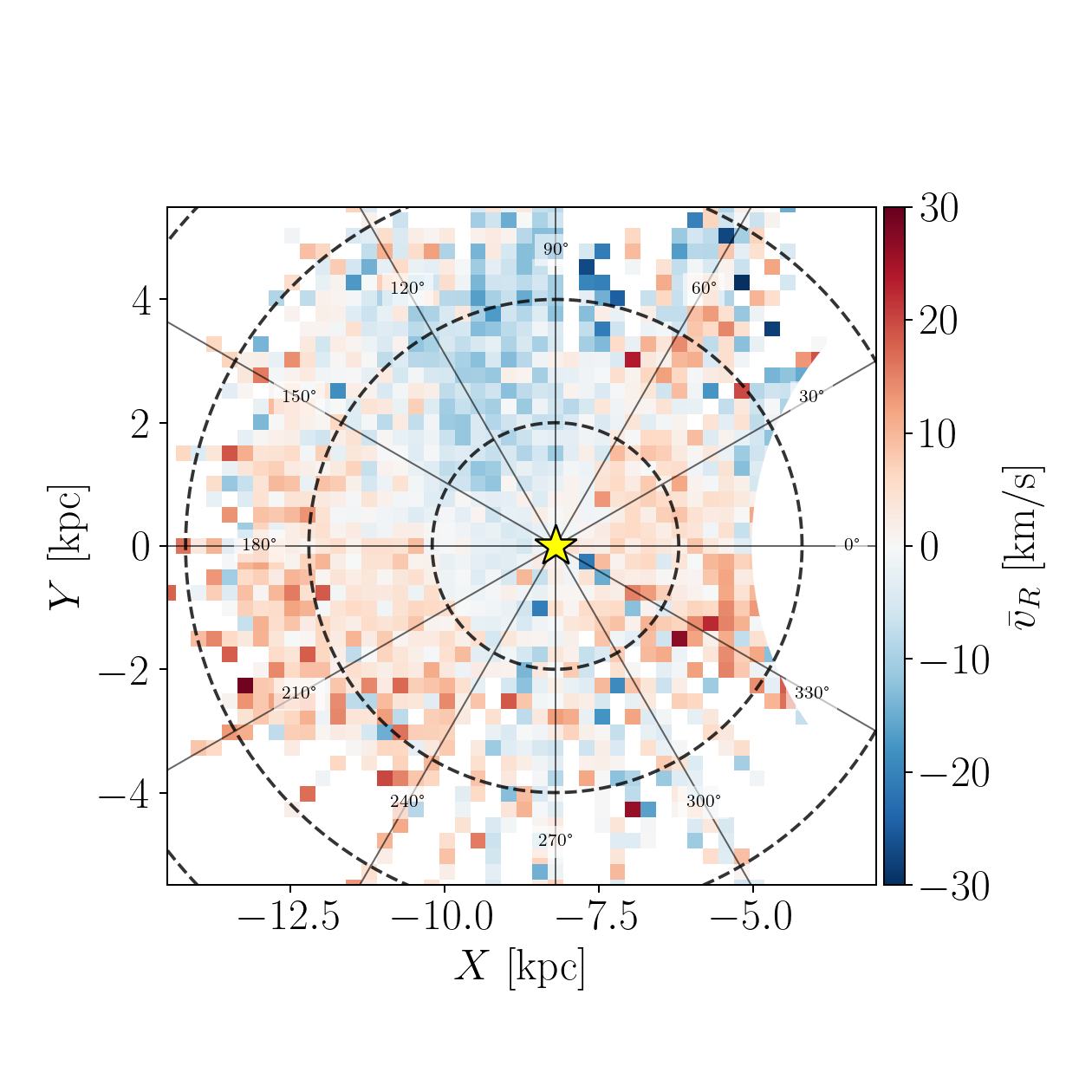}
    \caption{Distribution of red giant stars in the Milky Way disc from APOGEE DR17. The red giant stars are colour-coded by their average Galactocentric radial velocity  The location of the Sun is indicated by  the yellow star. The dashed black lines mark concentric circles at $\Delta R=2$ kpc.  The solid lines indicate constant Galactic longitudes. The region within 5 kpc from the Galactic centre is masked as the stellar motions in that region are influenced by the bar.}
    \label{fig:apogeeDR17}
\end{figure}

Another key question is how to relate the motions and orbits of young and old sources. Figure \ref{fig:apogeeDR17} shows an updated version  of Fig. 1 in \citet{Eilers2020}, generated by using APOGEE DR17 Red Giant Branch (RGB) stars selected as detailed in Appendix \ref{sec:appendixC}.  \citet{Eilers2020} interpreted the $\bar{v}_R$ feature visible in Fig. \ref{fig:apogeeDR17} as the kinematic signature of a logarithmic spiral arm.  This feature differs morphologically from that in Fig. \ref{fig:map-galcenrv}. Towards the Galactic centre, at $R_   {\mathrm{GC}} \lesssim 8$~kpc, the RGB sample shows positive $  \bar{v}_R$, while the HSS shows negative radial velocities. Moreover, the amplitude of the radial velocity asymmetries is lower by a factor of about 3. Instead, towards the outer Galaxy, some similarities are visible. For instance, in the region corresponding to the Perseus arm—between 2 and 4 kpc from the Sun and $Y > 0$~kpc—both samples show negative radial velocities, while for $Y < 0$~kpc, the velocities switch to positive values.   Similarly to the observed data, the radial velocity distribution of the young and old populations in the \texttt{aurigaia} mock data set is markedly different. Figure \ref{fig:auriga} (bottom) shows the density distribution (left) and radial velocities (right) of a population of giant stars extracted from the \texttt{aurigaia} catalogue using the same selection criteria described in Appendix \ref{sec:appendixC}.  The giant star population displays (a) a quadrupole feature associated with the Galactic bar, and (b) alternating inward and outward motions across the disc. While these motions are stronger than those observed in the data (Fig. \ref{fig:apogeeDR17}), the alternating pattern of positive and negative velocities is qualitatively similar.
The strength of the stellar response to spiral perturbations is expected to diminish with age. This attenuation arises from the increase in stellar velocity dispersion over time, which reduces the effectiveness of spiral arms in influencing older populations. As a result, a mixed stellar population responds differentially to the same spiral structure, depending on age and kinematic heating. This phenomenon, known as kinematic fractionation, was originally proposed to explain the vertical structure of X-shaped bulges~\citep{2017MNRAS.469.1587D}, but similar dynamics are at play across spiral arms as well~\citep{2018A&A...611L...2K}. This mechanism accounts for observed variations in stellar metallicity across spiral arms~\citep[e.g.][]{Poggio2022, Hawkins2023, Hackshaw2024}, and naturally explains the structure observed in the kinematic maps of young and old stellar populations in the data  and in the \texttt{aurigaia} mock data. Thus, young stars, though not necessarily formed within the prominent spiral arms, exhibit a stronger response to the spiral structure—as evidenced by their larger radial velocity amplitudes ($\bar{v}_R$)—in contrast to the smoother, more featureless kinematics of older stars. 

In general, spiral arms introduce small-scale radial velocity perturbations that are superimposed on the large-scale radial velocity pattern induced by the Milky Way bar. While the characteristic quadrupole pattern associated with the bar's influence inside its corotation radius ($<5$–$6$ kpc, see e.g. D23) is largely absent in Figs.\ref{fig:map-galcenrv} and~\ref{fig:compare_rv} (and similarly in the mock Auriga simulation; see Fig.\ref{fig:auriga}, top), the bar also generates non-circular motions at larger radii. The disc's response to the spiral structure and the Galactic bar has been explored by \citet[][see their Fig. 9]{2024arXiv241116866K} and
\citet[][M16a, M16b]{Monari2016a, Monari2016b}.
M16a found that stars located on the arms tend to move inward ($\bar{v}_R < 0$), while stars in the inter-arm regions move outward ($\bar{v}_R > 0$). M16b confirmed these findings and additionally showed that the Galactic bar dominates the horizontal motions ($v_R$ and $v{\phi}$). These results are in qualitative agreement with the features seen in Fig. \ref{fig:apogeeDR17}—cf. Fig. 1 in \citet{Monari2016a} and Figs. 4–8 in \citet{Monari2016b}—but they do not reproduce the $\bar{v}_R$ features shown in Fig. \ref{fig:map-galcenrv} for the hot star sample.
To isolate the localized radial velocity perturbations introduced by spiral arms, it is necessary to subtract the bar-induced bulk motion. This subtraction, however, requires assumptions regarding the 3D mass distribution of the bar and its pattern speed.

    \label{fig:R-vphi}

Modelling the response of populations of different ages to the spiral arm potential or to other perturbations is beyond the scope of this paper, and will be the focus of future works. The important point, at present, is that old and young stars exhibit remarkably different kinematic patterns, particularly within the solar circle, both in real and simulated data. These patterns are likely due to the effects on young stars of the spiral structure and the bar of our Galaxy.

\section{Summary and conclusions}\label{sec:conclusions}
In this study we present the first kinematic analysis of young OB stars in the Milky Way disc based on BOSS spectroscopic observations from SDSS-V, combined with astrometry from \textit{Gaia} DR3. 

We constructed a sample of spectroscopically confirmed hot stars (HSS, $T_{\mathrm{eff}} > 10,000$~K) with reliable radial velocities and distances, covering almost the entire Galactic plane and extending to distances up to $\sim$~5 kpc from the Sun. 

We derived the 3D motions of this sample  and we studied the distribution on the $X-Y$ plane of galactocentric radial velocities, $\bar{v}_R$. The HSS exhibit large-scale features in $v_R$, with amplitudes reaching $\pm 30 ~\mathrm{km~s^{-1}}$ . These velocity structures are spatially coherent over kiloparsec scales. 

We estimated the ages of all sources in the HSS with a Bayesian isochrone fitting procedure. These age estimates provide an indication of the youth of the sources more than an exact estimate. While the distribution of sources younger than 30 Myr closely follows the density enhancements of the entire sample, the observed kinematic patterns are not straightforwardly aligned with known spiral arms, and are not associated only to the young (< 30 Myr) or older sources (> 150 Myr).

We compared our results with mock catalogues generated from the \texttt{AURIGA} cosmological simulations \citep[][]{Grand2018} and with other observational studies based on different stellar samples. The radial velocity features observed in the data are observed in simulations as well, for sources younger than 500 Myr. Most of the observational studies that focus on the kinematics of the disc use samples of older sources. It is therefore not surprising that the features observed in the $\bar{v}_R$ distribution in the $X-Y$ plane \citep[e.g.][]{Eilers2020}  
for older sources, such as red giant branch stars, show different properties compared to those we observe in our sample.  

A plausible interpretation of the observed kinematic structures in the young stellar population is that they reflect a superposition of the effect of the Milky Way bar and spiral arms. A more detailed analysis and modelling is left for future work. This  will require a more complete sample (SDSS-V will continue observations until 2027, and the current data set amounts to around a third of the anticipated survey sample) as well as a model of Galactic structure and dynamics including populations of different ages.

\begin{acknowledgements}
We thank the referee for their constructive comments that helped improve the quality of this manuscript. \\

Funding for the Sloan Digital Sky Survey V has been provided by the Alfred P. Sloan Foundation, the Heising-Simons Foundation, the National Science Foundation, and the Participating Institutions. SDSS acknowledges support and resources from the Center for High-Performance Computing at the University of Utah. SDSS telescopes are located at Apache Point Observatory, funded by the Astrophysical Research Consortium and operated by New Mexico State University, and at Las Campanas Observatory, operated by the Carnegie Institution for Science. The SDSS web site is \url{www.sdss.org}.

SDSS is managed by the Astrophysical Research Consortium for the  Participating Institutions of the SDSS Collaboration, including the  Carnegie Institution for Science, Chilean National Time Allocation  Committee (CNTAC) ratified researchers, Caltech, the Gotham  Participation Group, Harvard University, Heidelberg University, The 
Flatiron Institute, The Johns Hopkins University, L'Ecole polytechnique  f\'{e}d\'{e}rale de Lausanne (EPFL), Leibniz-Institut f\"{u}r Astrophysik Potsdam (AIP), Max-Planck-Institut f\"{u}r Astronomie (MPIA Heidelberg), Max-Planck-Institut f\"{u}r Extraterrestrische Physik (MPE), Nanjing University, National Astronomical Observatories of China 
(NAOC), New Mexico State University, The Ohio State University, Pennsylvania State University, Smithsonian Astrophysical Observatory, Space Telescope Science Institute (STScI), the Stellar Astrophysics Participation Group, Universidad Nacional Aut\'{o}noma de M\'{e}xico, 
University of Arizona, University of Colorado Boulder, University of Illinois at Urbana-Champaign, University of Toronto, University of Utah, University of Virginia, Yale University, and Yunnan University. \\

The research leading to these results has (partially) received funding from the Research Foundation Flanders (FWO) under grant agreement G089422N, as well as from the BELgian federal Science Policy Office (BELSPO) through the PRODEX grant for PLATO.\\

This work has made use of data from the European Space Agency (ESA) mission {\it Gaia} (\url{https://www.cosmos.esa.int/gaia}), processed by the {\it Gaia} Data Processing and Analysis Consortium (DPAC,
\url{https://www.cosmos.esa.int/web/gaia/dpac/consortium}). Funding for the DPAC
has been provided by national institutions, in particular the institutions
participating in the {\it Gaia} Multilateral Agreement.\\

Software:Astropy, \citep{Astropy2013, Astropy2018, astropy2022}, matplotlib \citep{matplotlib}, numpy \citep{harris2020array}, scipy \citep{2020SciPy-NMeth}, topcat \citep{topcat2005}.
\end{acknowledgements}

\bibliographystyle{aa} 
\bibliography{bibliography.bib}

\appendix
\section{Comparison with D23}
\label{appendix:comparisonDrimmel}
\begin{figure}
    \centering
    \includegraphics[width = 0.4\textwidth]{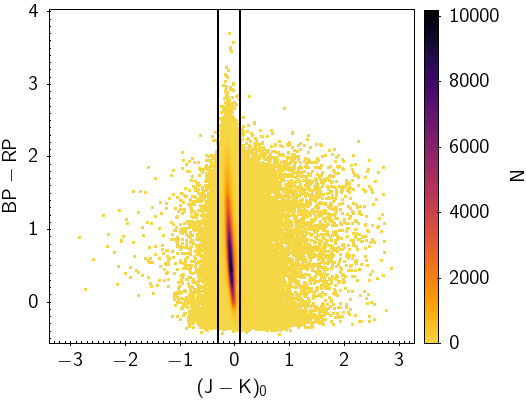}
     \includegraphics[width = 0.4\textwidth]{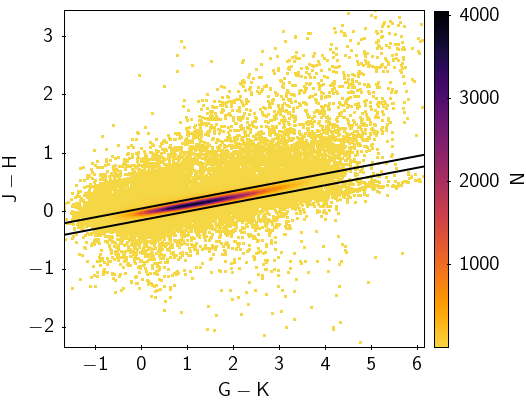}
    \includegraphics[width = 0.4\textwidth]{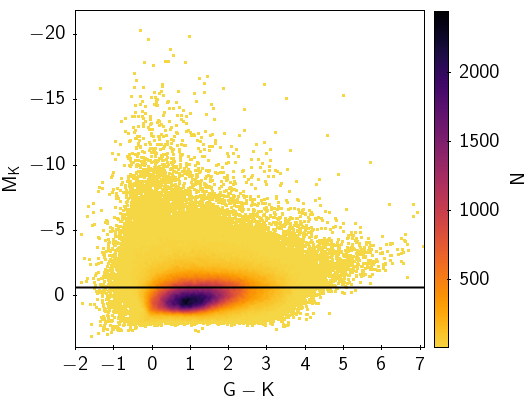}
     \caption{Distribution of the OB stars selected \cite{GaiaDrimmel2023} in the colour-colour and colour-magnitude diagrams used for the selection of the targets of the SDSS-V \texttt{mwm\_ob} program presented in \cite{Zari2021}. The solid black lines represent the selection criteria in \cite{Zari2021}.}
    \label{fig:cc and cmds}
\end{figure}

Figure \ref{fig:cc and cmds} shows the distribution of the D23 sample in the same colour-colour and colour-magnitude diagrams used to perform the target selection of the \texttt{mwm\_ob} program of the SDSS-V survey. With regards to the colour-colour diagrams (top and middle panel), most of the sources selected in D23  fall within the selection criteria of Z21. The bottom panel shows that the absolute magnitude distribution in the 2MASS $K_s$ band ($M_\mathrm{K}$) extend to $M_\mathrm{K_s} \sim 2$~mag, while, as mentioned in the main text, our selection included only stars brighter than $M_\mathrm{K_s} = -0.6$~mag.
Unfortunately, extending  the selection criteria in Z21 by including all stars brighter than $M_\mathrm{K_s} = 2$~mag ($\varpi < 10^{(10 - k + 2)/5}$) would result in a much larger number of sources than those observable within the duration of the SDSS-V survey.

\begin{figure*}
    \includegraphics[width = 0.49\textwidth]{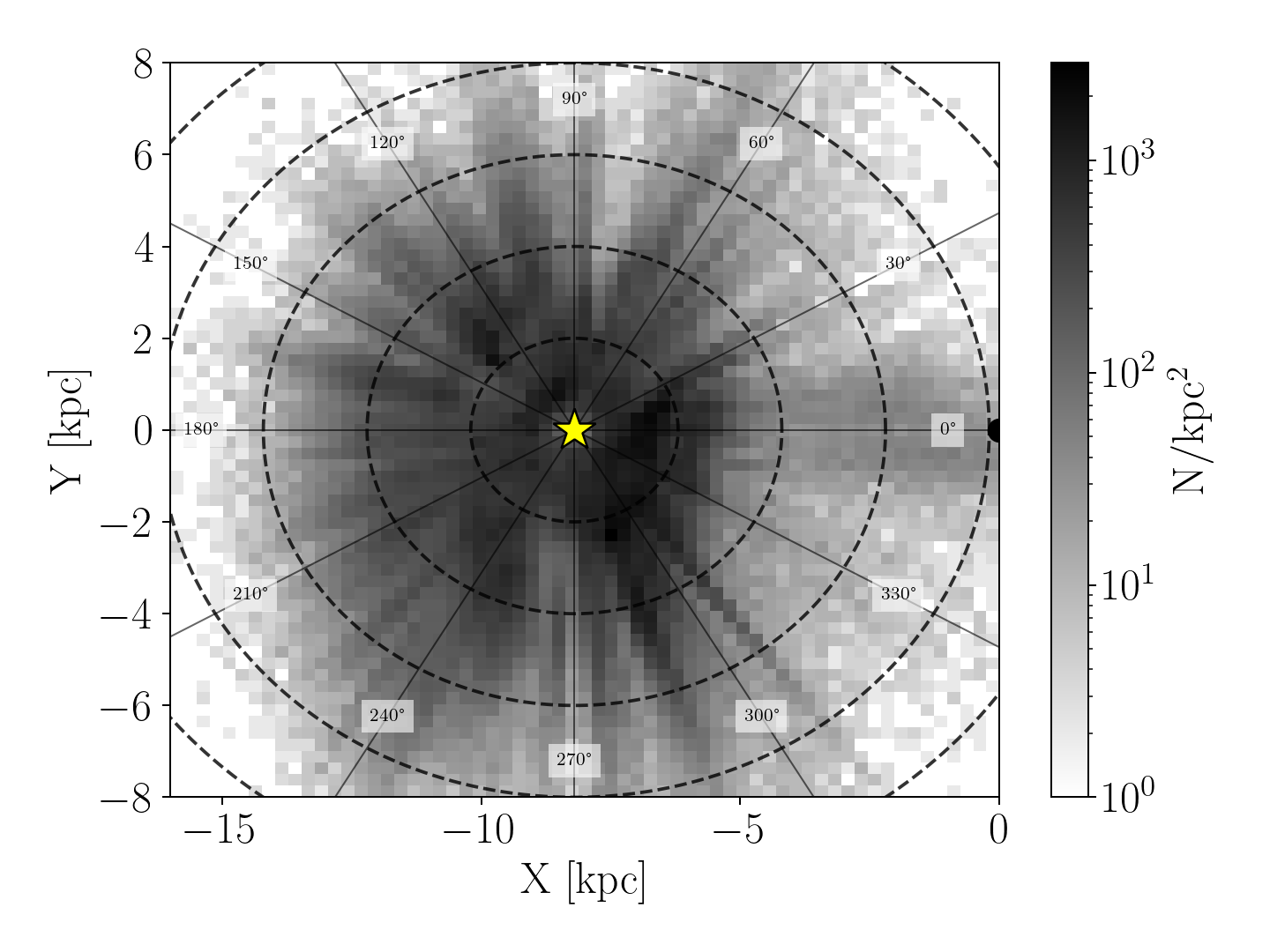}
    \includegraphics[width = 0.49\textwidth]{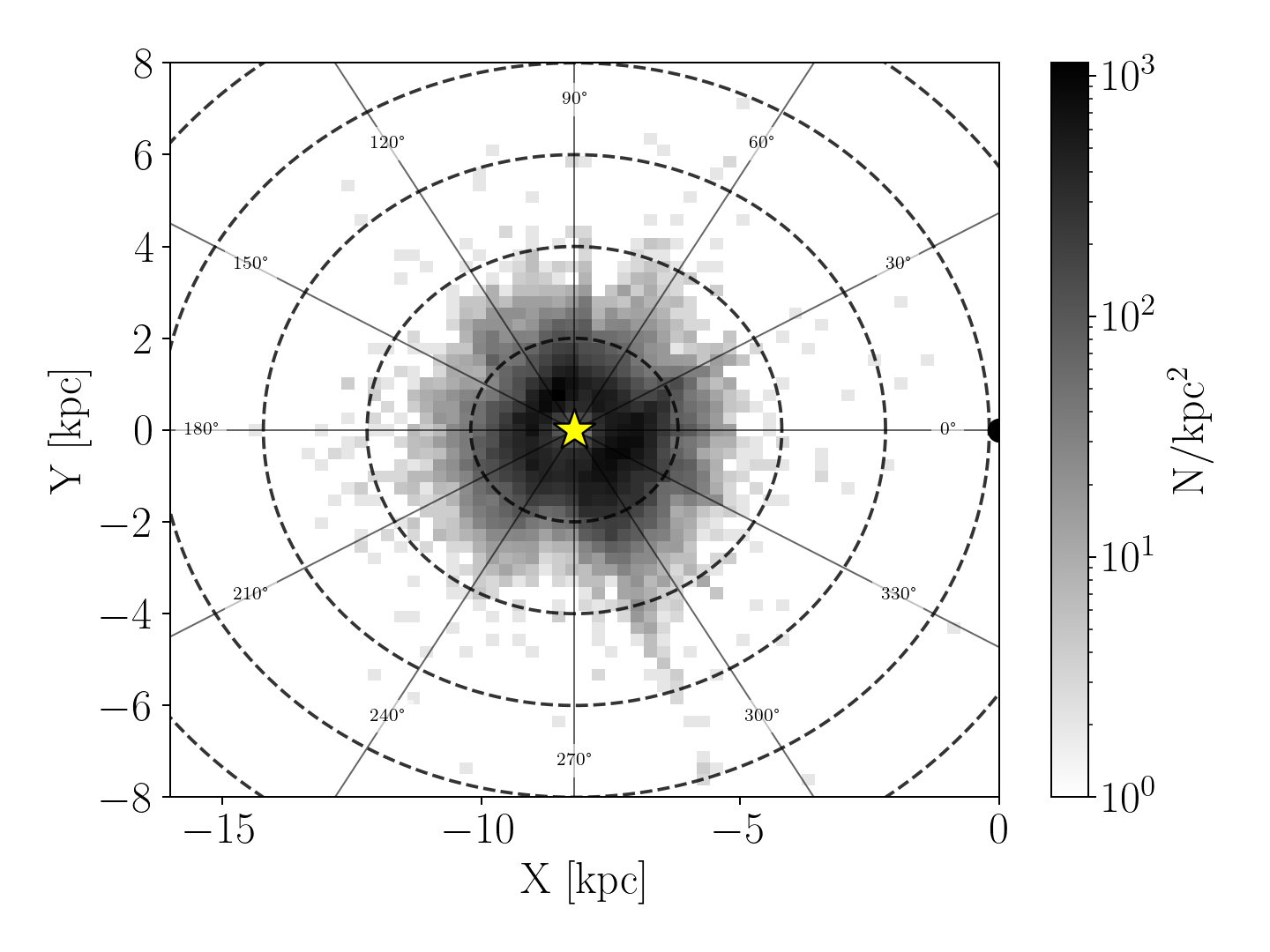}
\caption{Distribution of the sample presented in D23 in the Galactic plane. The left panel shows the entire sample, the right panel shows only the sources with \textit{Gaia} DR3 line-of-sight velocities. The yellow star represents the Sun's position. The dashed black circles have radii $R = 2, 4, 6, 8, 10$~kpc. The solid lines indicate constant Galactic longitudes.}
\label{fig:drimmel_plane}
\end{figure*}

Figure \ref{fig:drimmel_plane}
shows the distribution in the Galactic plane of the D23 sample with (right) and without (left) \textit{Gaia} DR3 line-of-sight velocities (cfr. \ref{fig:map-galcenrv}).
While the full \textit{Gaia} DR3 sample (left) covers a similar -- if not larger -- area of the disc with respect to the SDSS-V \texttt{mwm\_ob}/HSS, the \textit{Gaia} DR3 line-of-sight velocity sample covers a much smaller one.  This is likely due to the selection function of the \textit{Gaia} line-of-sight velocity sample. Indeed,  line-of-sight velocities are derived for $T_{\mathrm{eff}}$ between 6900-14500~K   and are limited to stars with $G_{\mathrm{RVS}} < 12$~mag \citep{Blomme2023}. 

\section{Maps of velocity dispersion}
\begin{figure*}
    \centering
    \includegraphics[width=0.49\linewidth]{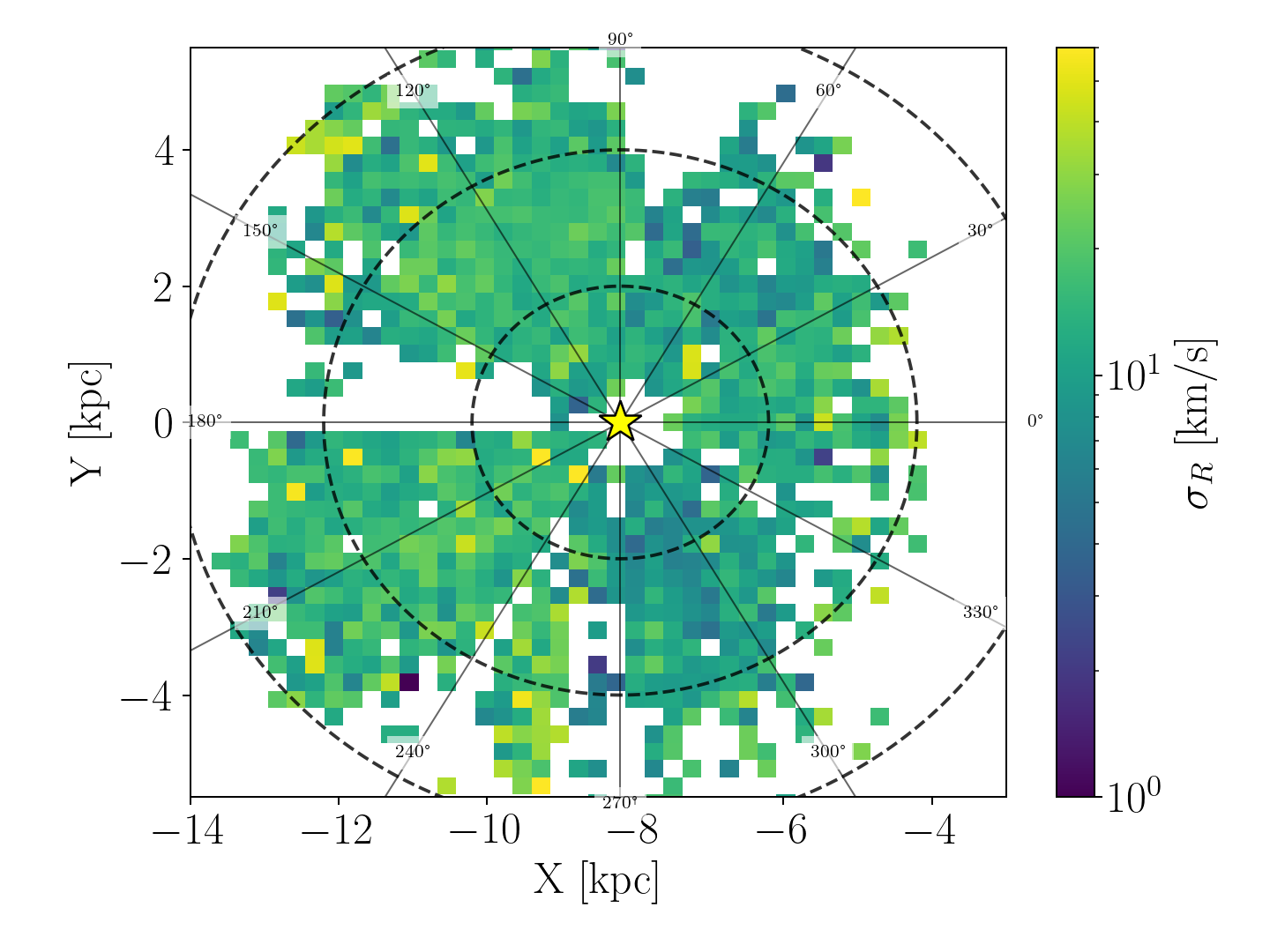}
    \includegraphics[width=0.49\linewidth]{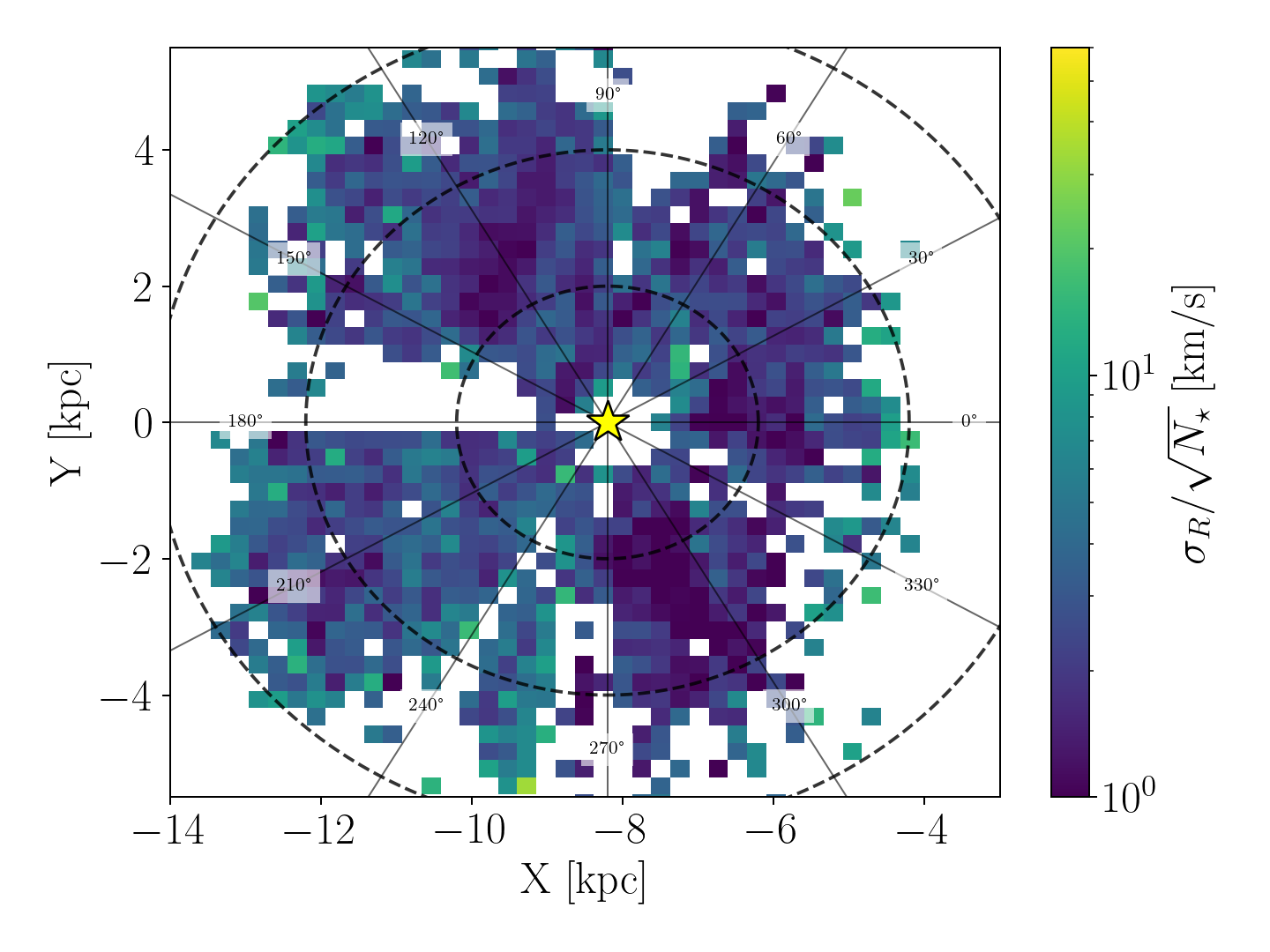}\\
    \includegraphics[width=0.49\linewidth]{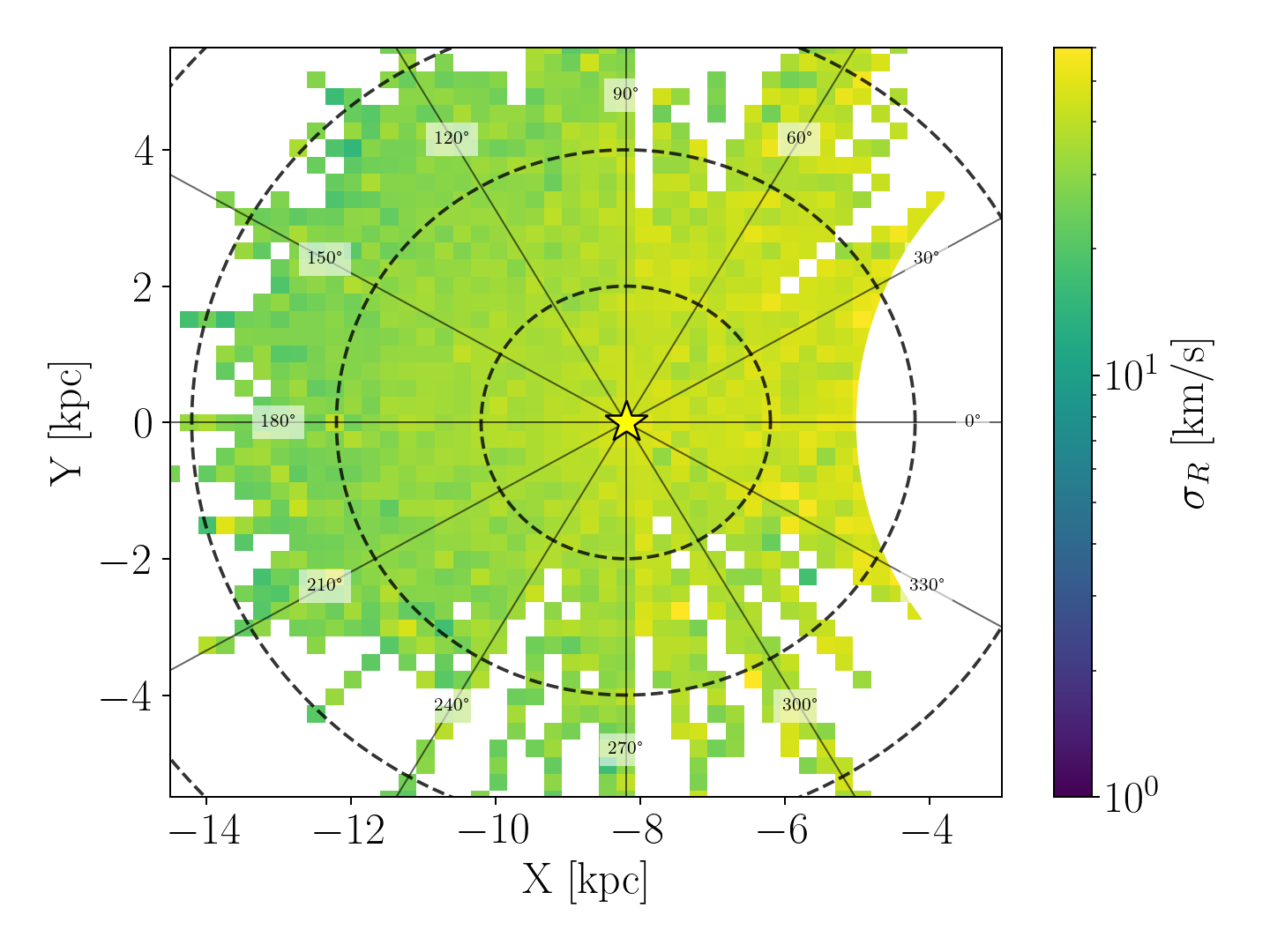}
    \includegraphics[width=0.49\linewidth]{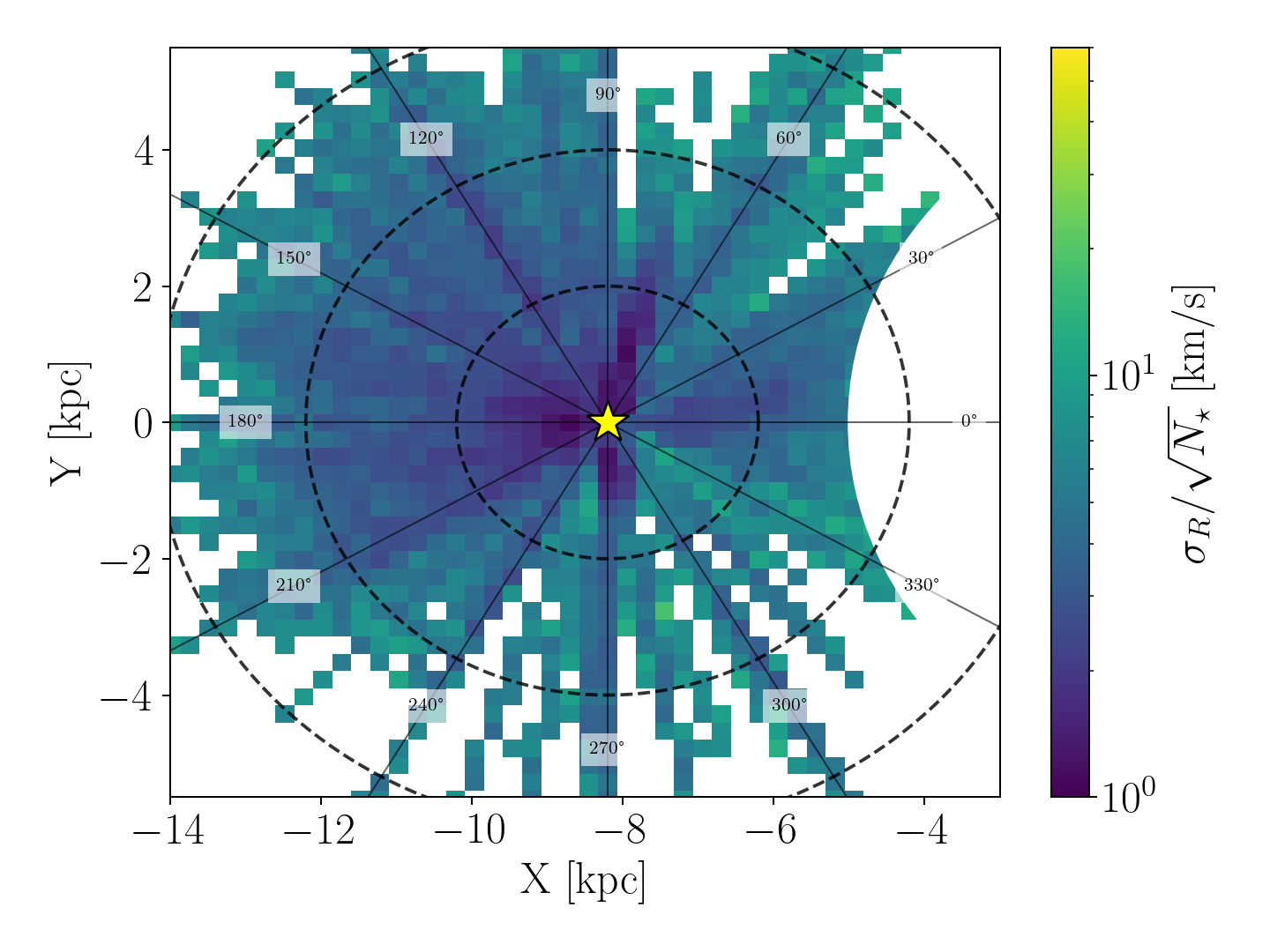}    
    \caption{Velocity dispersion (left) and velocity uncertainty (right) maps for the Hot Star Sample (top) and for the APOGEE DR17 sample (bottom). The colorbars extend from 1 km/s to 60 km/s in all panels. The yellow star marks the location of the Sun, in X,Y=(-8.2,0). The dashed circles have radii separated by $\Delta R = 2$~kpc.}
    \label{fig:vR_error}
\end{figure*}

Figure \ref{fig:vR_error} (left) shows the maps of the velocity dispersion $\sigma_R$ derived by using Eq. \ref{eq:likelihood} in Section \ref{section:3D pos and kin} for the HSS (top), and in Section \ref{sec:comparison_kin} for the APOGEE DR17 RGB sample (bottom, see Section \label{sec:appendixC}). 
The velocity dispersion of the hot star sample is usually between 10-20 km/s. This large value should not excessively surprise the reader, as the spatial scale of the map is large (the spaxel size is 250$\times$250 pc). We note that the velocity dispersion of the HSS is remarkably smaller than the velocity dispersion of the APOGEE DR17 sample (see Fig. \ref{fig:vR_error} bottom left). This is expected, as the kinematic properties of the two samples are dramatically different, and older populations have larger velocity dispersion.

We estimated the velocity uncertainty on $\bar{v}_R$ as $\sigma_R/\sqrt{N_{\star}}$, where $N_{\star}$ is the number of stars in a spaxel. Figure  \ref{fig:vR_error} (right) shows the maps of the radial velocity uncertainty $\sigma_R/\sqrt{N_{\star}}$ for the HSS (top) and the APOGEE DR17 RGB sample (bottom). 

\section{Simulation of a population of stars with 50\% binary fraction}\label{appendix:vlos}
\begin{figure}
    \centering
    \includegraphics[width=\linewidth]{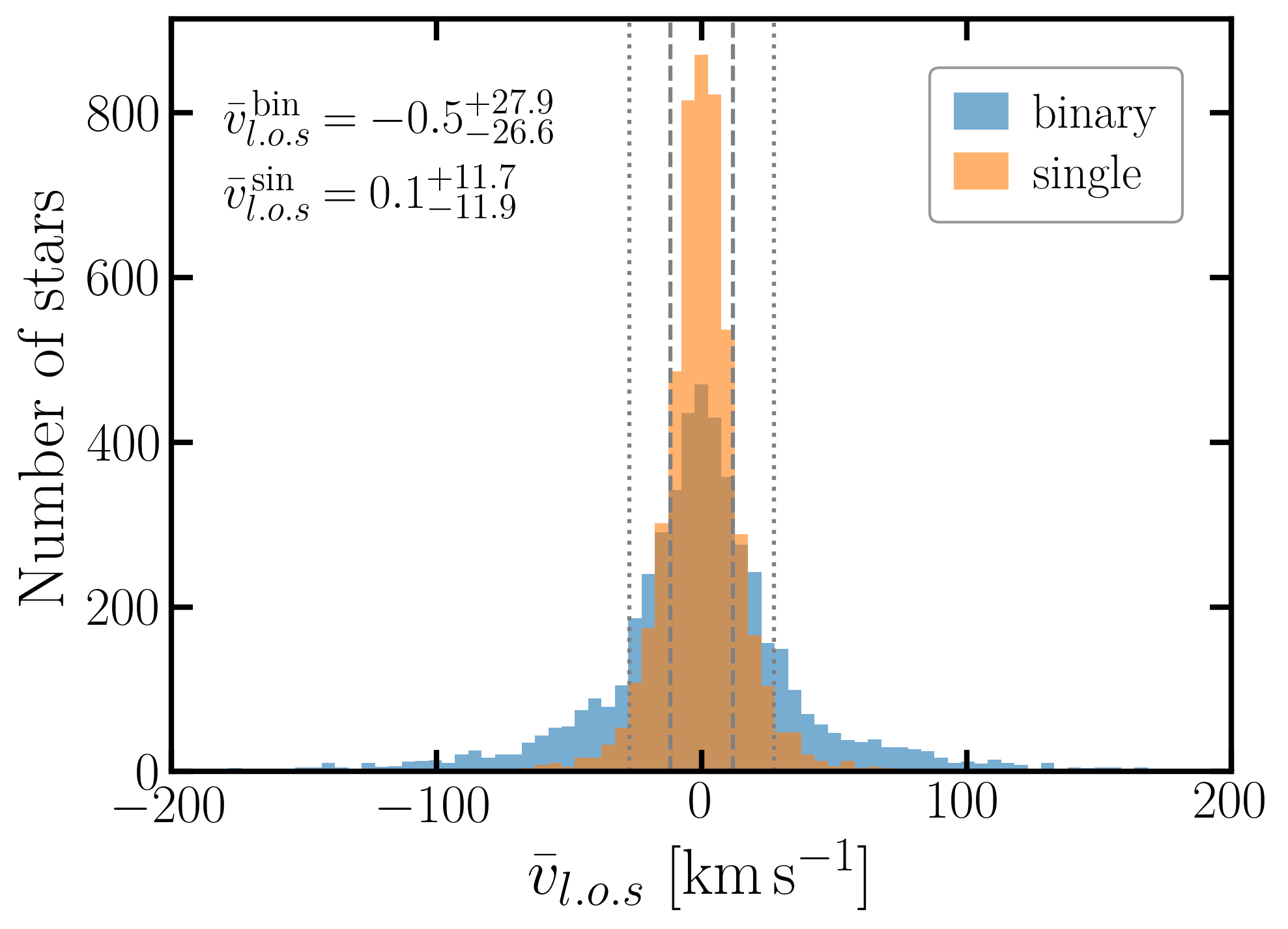}
    \caption{Distribution of average line-of-sight velocities for a simulated population of stars with 50\% binary fraction, observed with the same cadence as our sample of Hot Stars, and with the same $v_{l.o.s.}$ errors. Vertical lines mark the 16th/84th percentile for single (dashed) and binary (dotted) stars. Annotated labels report the mean and corresponding uncertainty intervals derived from the 16th and 84th percentiles.}
    \label{fig:sim_vlos}
\end{figure}

As explained in Section \ref{sec:vel}, we verified that the average is a good estimator of the centre of mass motion of our sample by simulating a population of stars with 50\% binary fraction, observed at the same MJD's and with the same $v_{l.o.s.}$ errors as our sample. Figure \ref{fig:sim_vlos} 
shows  histogram distributions of the average velocities for simulated (blue) and single (orange) stars.
For single stars, the dispersion around the average value roughly coincides with the measurement errors. For binary stars, the spread is larger. The main contribution to the spread is due to binary stars that have only a single-epoch observation. 
We repeated our analysis by including only stars observed at $\ge 2$ epochs, and we did not find any significant difference on our main results (Figs. \ref{fig:map-galcenrv} and \ref{fig:plane_ages}).

\section{Selection of the APOGEE DR17 sample of disc stars}\label{sec:appendixC}
\begin{figure}
    \centering
    \includegraphics[width=0.5\textwidth]{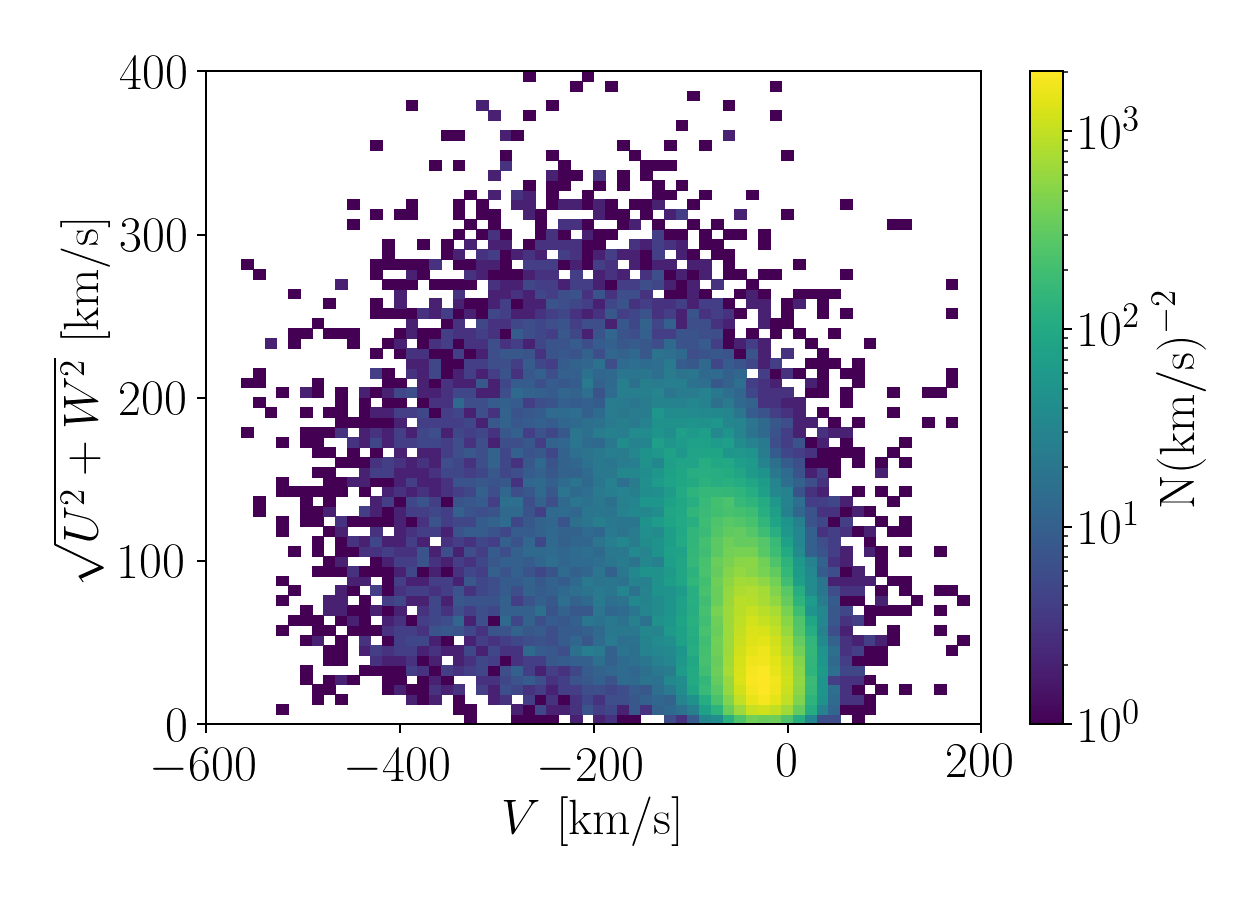}
    \caption{Toomre diagram for the APOGEE DR17 \texttt{astroNN} sample described in Sec. \ref{sec:appendixC}. }
    \label{fig:toomre}
\end{figure}

To compare the motions of young massive stars and old disc sources in Sec. \ref{sec:discussion} we made use of the \texttt{astroNN} value-added catalogue \citep[][]{Leung2019} of abundances, distances, and ages for APOGEE DR17 sources\footnote{The catalogue can be downloaded at this link: \url{https://data.sdss.org/sas/dr17/env/APOGEE_ASTRO_NN/}}.  \texttt{astroNN} is an open-source Python package developed for the neural network trained on the APOGEE data and is designed to be a general package for deep learning in astronomy. The
APOGEE-\texttt{astroNN} catalogue contains results from applying \texttt{astroNN} neural nets on APOGEE DR17 spectra to infer stellar parameters, abundances trained with ASPCAP DR17 \citep[][Holtzman et al., in prep]{Shetrone2015, Garcia2016, Smith2021}, distances retrained with Gaia eDR3 \citep[][]{GaiaEDR32021} from \cite{Leung2019b}, and ages trained with APOKASC-2 \citep[][]{Mackereth2019} in combination with low-metallicity asteroseismic ages \citep[][]{Montalban2021}.
Following \cite{Hackshaw2024},  we employed the
following cuts to select disc stars:
\begin{enumerate}
    \item We selected stars in the plane by requiring $|b| < 10$ deg.
    \item We used the \texttt{astroNN} distance and distance errors to select stars with distance errors <30\%.
    \item To select red giants, we restricted our sample to stars in the effective temperature range $3500 < \mathrm{Teff} < 5000$ K, and with $\log g < 3.6$ dex.
    \item Finally, we select stars following disc kinematics by using the Toomre diagram (see Fig. \ref{fig:toomre}). In particular we selected stars with $-600 < V < 200$~km/s and $\sqrt{U^2 + W^2} < 400$~km/s.
    This is a broad selection, that nevertheless allows to remove outliers.
\end{enumerate}
\texttt{astroNN} provides pre-computed positions, velocities, and actions. Positions and velocities were computed assuming the Sun is located at 8.125 kpc from the Galactic center \citep[][]{gravity2019}, 20.8 pc above the Galactic midplane   \citep{bennett2019}, and has radial, rotational, and vertical velocities of -11.1, 242, and 7.25 km/s. Actions were computed by integrating the stellar orbits with the Gala code \citep{Price-Whelan2017}. The mass-model of the
Milky Way used for the gala.dynamics.orbit function
was the MilkyWayPotential2022, which has been fitted
to the rotation curve in \cite{Eilers2020} and
incorporates the phase-space spiral in the solar neighborhood set by E. Darragh-Ford et al. (2023).

\end{document}